\documentclass{IEEEtran}
\def\BibTeX{{\rm B\kern-.05em{\sc i\kern-.025em b}\kern-.08em
		    T\kern-.1667em\lower.7ex\hbox{E}\kern-.125emX}}

\usepackage{hyperref}
\hypersetup{hidelinks=true}
\usepackage{textcomp}

\usepackage{amsmath,amssymb,amsfonts}
\usepackage{mathabx}
\usepackage{algorithmic}
\usepackage{graphicx}
\usepackage{textcomp}
\usepackage[utf8]{inputenc}
\usepackage[T1]{fontenc}
\usepackage{amsmath}
\usepackage{amsfonts}
\usepackage{amssymb}
\usepackage{balance}
\usepackage{graphicx}
\usepackage[rightcaption]{sidecap}
\usepackage{import}
\usepackage{textcomp}
\usepackage[dvipsnames]{xcolor}
\usepackage{bbm}
\usepackage{slashed}
\usepackage{caption}
\usepackage{subcaption}
\usepackage{lipsum}
\usepackage{enumitem}
\usepackage{tabularx}
\usepackage{xcolor}
\usepackage[ruled,linesnumbered]{algorithm2e}
\usepackage{leftindex}
\usepackage{mathtools}
\usepackage{booktabs}
\usepackage{makecell}
\usepackage[noadjust]{cite}
\usepackage{multirow}

\usepackage{easyReview}
\usepackage{bm}

\let\underbrace\LaTeXunderbrace

\SetKwComment{Comment}{/* }{ */}
\SetKwInOut{Input}{Input}
\SetKwInOut{Output}{Output}
\SetKwInOut{Parameters}{Parameters}

\newcommand{\bbR}{\mathbb{R}}

\newcommand{\calB}{\mathcal{B}}
\newcommand{\calC}{\mathcal{C}}
\newcommand{\calD}{\mathcal{D}}

\newcommand{\calF}{\mathcal{F}}

\newcommand{\calH}{\mathcal{H}}

\newcommand{\calK}{\mathcal{K}}
\newcommand{\calL}{\mathcal{L}}
\newcommand{\calM}{\mathcal{M}}
\newcommand{\calN}{\mathcal{N}}
\newcommand{\calO}{\mathcal{O}}
\newcommand{\calP}{\mathcal{P}}

\newcommand{\calR}{\mathcal{R}}
\newcommand{\calS}{\mathcal{S}}

\newcommand{\calU}{\mathcal{U}}
\newcommand{\calV}{\mathcal{V}}

\newcommand{\calX}{\mathcal{X}}

\usepackage{amsthm}

\theoremstyle{definition}
\newtheorem{assumption}{Assumption}
\newtheorem{theorem}{Theorem}
\newtheorem{lemma}[theorem]{Lemma}
\newtheorem{proposition}[theorem]{Proposition}
\newtheorem{corollary}[theorem]{Corollary} 
\newtheorem{definition}{Definition}

\newtheorem{example}{Example}

\theoremstyle{remark}
\newtheorem{remark}{Remark}

\DeclareMathOperator*{\argmax}{arg\,max}

\begin{document}
\title{Leveraging Equivariances and Symmetries in the Control Barrier Function Synthesis}
\author{Adrian Wiltz, and Dimos V. Dimarogonas
\thanks{This work was supported by the ERC Consolidator Grant LEAFHOUND, the Horizon Europe EIC project SymAware (101070802), the Swedish Research Council, and the Knut and Alice Wallenberg Foundation.}
\thanks{The authors are with the Division of Decision and Control Systems, KTH Royal Institute of Technology, SE-100 44 Stockholm, Sweden {\tt\small \{wiltz,dimos\}@kth.se}.}}

\maketitle
\begin{abstract}
The synthesis of Control Barrier Functions (CBFs) often involves demanding computations or a meticulous construction. However, structural properties of the system dynamics and constraints have the potential to mitigate these challenges. In this paper, we explore how equivariances in the dynamics, loosely speaking a form of symmetry, can be leveraged in the CBF synthesis. Although CBFs are generally not inherently symmetric, we show how equivariances in the dynamics and symmetries in the constraints induce symmetries in CBFs derived through reachability analysis. This insight allows us to infer their CBF values across the entire domain from their values on a subset, leading to significant computational savings. Interestingly, equivariances can be even leveraged to the CBF synthesis for non-symmetric constraints. Specifically, we show how a partially known CBF can be leveraged together with equivariances to construct a CBF for various new constraints. Throughout the paper, we provide examples illustrating the theoretical findings. Furthermore, a numerical study investigates the computational gains from invoking equivariances into the CBF synthesis.
\end{abstract}



\section{Introduction}

The dynamic properties of a system can be effectively and precisely characterized using Lyapunov-like functions. They capture stability and convergence properties, and allow to identify forward invariant sets. Certain notions of Lyapunov-like functions have been specifically introduced for controller synthesis. As such, Control Lyapunov Functions (CLF)~\cite{Sontag1989} are used for stabilization, while their control-theoretic counterpart for enforcing forward set-invariance are Control Barrier Functions (CBF)~\cite{Wieland2007,Ames2017}. The focus in this paper is on CBFs. Since they provide a control-oriented characterization of a system's dynamic capabilities with respect to state constraints, they have become a widely used design tool in numerous fields of control such as robotics and vehicle control, coordination and spatio-temporal logic constraints, and more \cite{Lindemann2019,Wiltz2022a,Nguyen2023,Frauenfelder2023,Molnar2025,Wong2025}. While the controller design is straight-forward once a CBF has been found, the process of finding a CBF generally involves a meticulous construction or computationally demanding methods. 

In this paper, we exploit the structural properties of system dynamics and constraints to reduce the complexity of CBF synthesis. In particular, we focus on equivariant systems, which --- loosely speaking --- exhibit some type of symmetry in their dynamics. Examples of such systems include vehicles and mobile robots, attitude kinematics and robotic manipulators~\cite{Selig2007,Apraez2025,Mahony2022}.

The fundamental implications of equivariances for dynamical systems have been investigated in~\cite{Field1970,Field1980}. Notably, one solution to an equivariant system allows to characterize further solutions of the same system. Equivariances thereby represent an important system theoretic property that found application across diverse fields such as bifurcation theory and fluid dynamics~\cite{Crawford1991,Chossat2000}, systems biology~\cite{Shoval2011,Sontag2017}, and more recently, in the design of convolutional neural networks for image and shape recognition~\cite{Lenc2015,Cohen2018,Finzi2020}. Within control engineering, equivariances play a significant role in observer design~\cite{Mahony2020,Mahony2022,Goor2023}, have been incorporated into data-driven control methods~\cite{Sinha2020,Bousias2025}, and can be found in many robotic systems~\cite{Selig2007,Apraez2025}. Despite their important role in a broad range of applications, the use of equivariances in the controller synthesis has only gained momentum in recent years. 

In the synthesis of Lyapunov-like functions, and CBFs in particular, equivariances have received little attention so far. The construction of CBFs offers much freedom in designing a value function that satisfies a certain gradient condition under the provided system dynamics and constraints. While the symmetry of a CBF is typically not inherited from the CBF design problem, some dynamics and constraints still permit the construction of a symmetric CBF. This symmetry can be exploited during synthesis, as it allows CBF values to be determined from a known value at another state as we show this in the paper. However, even beyond symmetric constraints, equivariances can be leveraged to synthesize CBFs for new constraints based on an existing CBF, which we show as well. 

To the best of our knowledge, the only other work considering equivariances in the CBF synthesis so far is~\cite{Bousias2025}, which utilizes equivariances for the learning-based synthesis of CBFs in multi-agent systems with limited data available. It is however confined to symmetric constraints. 

The contributions of this paper are twofold. At first, we show how equivariances in the dynamics and symmetries in the constraints induce symmetries in CBFs derived via Hamilton-Jacobi reachability. In particular, we build on the reachability-based CBF construction proposed in~\cite{Wiltz2025b}, which allows for pointwise computation and extends to certain time-varying constraints. The results allow us to infer CBF values across the entire domain from values on a subset. 

Secondly, we extend this approach to a broader class of constraints, including those that are not symmetric. Specifically, we show how a partially known CBF, combined with the system's equivariances, can be used to infer CBFs for new, potentially non-symmetric constraints. This second method is agnostic to how the original CBF was synthesized and can be applied together with any synthesis technique. Provided that the system dynamics possess equivariances, our contributions lead to a significantly reduced computational effort during CBF synthesis. Throughout the paper, we support our theoretic results with numeric examples that illustrate their application. 

The outline of the paper is as follows. Section~\ref{sec:background} provides some background. Section~\ref{sec:equivarainces for symmetric constraints} leverages equivariances and symmetries in the synthesis of reachability-based CBFs; case studies on the application of the results are given in Section~\ref{sec:case studies}. Section~\ref{sec:beyond the symmetric case} extends the results beyond symmetric constraints. Section~\ref{sec:numeric examples} investigates the computational gains from equivariances and demonstrates the synthesis of CBFs for various constraints from a single partially known CBF. Section~\ref{sec:conclusion} draws a conclusion.

\emph{Notation:} Let $ \calX\subseteq\bbR^{n} $, $ x\in\bbR^{n} $. Sets are denoted by calligraphic uppercase letters, trajectories $ \bm{x}:\bbR \rightarrow \calX $ by bold lowercase. The set of trajectories on $ [t_{1},t_{2}] $ is $ \bm{\calX}_{[t_{1},t_{2}]} $, abbreviated as $ \bm{\calX} $ when clear from context. Complement, boundary and interior of $ \calX $ are denoted $ \calX^{c} $, $ \partial\calX $ and $ \text{Int}(\calX) $; the Euclidean norm by $ ||\cdot|| $. A ball of radius $ r $ around $ x_{0} $ is $ \calB_{r}(x_{0}) \coloneq \{x \, | \, ||x_{0}-x||< r\} $. For $ \calX_{1},\calX_{2}\subseteq\bbR^{n} $, the Minkowski sum is $ \calX_{1}\oplus\calX_{2} = \{x_{1}+x_{2} \, | \, x_{1}\in\calX_{1}, \, x_{2}\in\calX_{2}\} $, and the Pontryagin difference $ \calX_{1} \ominus \calX_{2} \coloneq \{x_{1} \in \bbR^{n} \, | \, x_{1} + x_{2} \in\calX_{1}, \; \forall x_{2}\in\calX_{2} \} $. A class~$ \calK $ function is a continuous, strictly increasing function $ \alpha: \bbR_{\geq0} \rightarrow \bbR_{\geq 0} $ with $ \alpha(0) = 0 $; if $ \alpha: \bbR\rightarrow\bbR $, then it is an extended class~$ \calK_{e} $ function. Furthermore, let $ g:\bbR^{n}\rightarrow\bbR^{m} $ and $ M\in\bbR^{n\times m} $. The image of set $ \calX $ is $ g(\calX)\coloneq\{g(x): x\in\calX\}$ and $ M\,\calX\coloneq\{Mx: x\in\calX\} $. The sign function is $ \text{sgn}(x) = 1 $ if $ x>0 $, $ 0 $ if $ x=0 $, and $ -1 $ otherwise. The scalar product is sometimes denoted by $ \langle\cdot,\cdot\rangle $. The standard unit vector is $ \bm{e}_{i} $ with 1 in the $ i $-th position; identity and zero matrices are $ \bm{I} $, $ \bm{I}_{n} $, $ \bm{0} $ and $ \bm{0}_{n} $. For $ p\in\bbR^{n} $, $ \text{diag}(p) $ denotes the diagonal matrix with entries of $ p $. The column span of $ M $ is $ \text{span}(M) $; its $ ij $-th entry and the $ j $-th column are $ [M]_{ij} $ and $ [M]_{\ast j} $. A property holds almost everywhere (a.e.) if it holds everywhere except on a measure-zero set.


\section{Background}
\label{sec:background}

Let us consider the state constraint
\begin{align}
	\label{eq:state constraint}
	x\in\calH\coloneq\{x\in\bbR^{n} \,|\, h(x)\geq 0\},
\end{align}
where $ h:\bbR^{n}\rightarrow\bbR $ is Lipschitz continuous. 
The state is governed by the dynamic system
\begin{align}
	\label{eq:dynamics}
	\dot{x}=f(x,u), \qquad x(0)=x_0,
\end{align}
where $ x, x_{0}\in\bbR^{n} $, $ u\in\bbR^{m} $, and $ f: \bbR^{n}\times\calU \rightarrow \bbR^{n} $ is Lipschitz continuous in both of its arguments. For an almost everywhere continuous input trajectory $ \bm{u}: \bbR_{\geq 0} \rightarrow \bm{\calU} $, the solution to~\eqref{eq:dynamics} up to some time~$ T $ is denoted by $ \bm{\varphi}: [0,T]\rightarrow\bbR^{n} $. It is given as $ \bm{\varphi}(t;x_{0},\bm{u}) \coloneq x_{0} + \int_{0}^{T} f(\bm{x}(\tau),\bm{u}(\tau)) d\tau $; its forward completeness is assumed. A set $ \calS\subseteq\bbR^{n} $ is called \emph{forward control invariant} with respect to dynamics~\eqref{eq:dynamics} if there exist trajectories $ \bm{u}\in\bm{\calU}_{[0,\infty]} $ such that $ \bm{\varphi}(t;x_{0},\bm{u})\in\calS $ for all $ t\geq0 $. Furthermore, \eqref{eq:dynamics} is said to be \emph{controllable} on $ \calS $ \cite{Hermann1977}  if $ \calS \subseteq \bigcup_{t\in[0,\infty)} \calR_{t}(x_{0}) $ for all $ x_{0}\in\calS $ where $ \calR_{T}(x_{0}) \coloneq \{x_{1} \, | \, \exists \bm{u}\in\bm{\calU}_{[0,T]}: \bm{\varphi}(T; x_{0},\bm{u}) = x_{1} \} $ is the set of \emph{$ T $-reachable states}. At last, a state $ x_{0}\in\bbR^{n} $ is said to be \emph{viable} if there exists $ \bm{u}:\bbR_{\geq 0} \rightarrow \calU $ such that $ \bm{\varphi}(t;x_{0},\bm{u})\in\calH $ for all $ t\geq0 $.

\subsection{Control Barrier Functions in the Dini Sense}

CBFs are often defined as differentiable functions, which can be limiting. Similarly to CLFs~\cite{Brockett1983,Clarke2011}, not all systems admit a smooth CBF. Moreover, requiring differentiability can complicate the synthesis~\cite{Glotfelter2020,Marley2024,Charitidou2023}. Instead, we consider the more general notion of CBFs in the Dini sense analogously to CLFs in the Dini sense~\cite{Clarke2011}.

\begin{definition}[CBF in the Dini Sense~\cite{Wiltz2025b}]
	\label{def:cbf dini}
	Let $ \mathcal{D} \subseteq \mathbb{R}^n $ and let $ b: \mathbb{R}^n \to \mathbb{R} $ be a locally Lipschitz continuous function. Define the set $ \mathcal{C} \coloneqq \{ x \in \mathbb{R}^n \mid b(x) \geq 0 \} $ and suppose $ \mathcal{C} $ is compact and satisfies $ \mathcal{C} \subseteq \mathcal{D} $. Then, $ b $ is called a \emph{Control Barrier Function (CBF) in the Dini sense} on $ \mathcal{D} $ with respect to dynamics~\eqref{eq:dynamics} if there exists an extended class $ \mathcal{K}_e $ function~$ \alpha $ such that for all $ x \in \mathcal{D} $,
	\begin{align}
		\label{eq:def cbf dini}
		\sup_{u \in \mathcal{U}} \{ db(x; f(x,u)) \} \geq -\alpha(b(x)),
	\end{align}
	where the Dini derivative of a locally Lipschitz function $ \phi: \mathbb{R}^n \to \mathbb{R} $ at $ x $ in the direction $ v \in \mathbb{R}^n $ is given by
	\begin{align*}
		d\phi(x; v) \coloneqq \liminf_{\sigma \downarrow 0} \frac{\phi(x + \sigma v) - \phi(x)}{\sigma}.
	\end{align*}
\end{definition}

The result on the controlled forward invariance of $ \calC $ is analogous to the differentiable case.

\begin{theorem}[\cite{Wiltz2025b}, Thm.~1]
	\label{thm:forward_invariance}
	Let $ \bm{u} \in \bm{\mathcal{U}}_{[0,T]} $ be continuous a.e., and let $ \bm{x}(t) \coloneqq \bm{\varphi}(t; x_0, \bm{u}) $ denote the corresponding state trajectory of~\eqref{eq:dynamics} starting in $ x_0 \in \mathcal{C} $. If $ db(\bm{x}(t);f(\bm{x}(t),\bm{u}(t)))\geq-\alpha(b(\bm{x}(t))) $ for all $ t\in[0,T] $, then set $ \mathcal{C} $ is forward invariant.
\end{theorem}

\subsection{Reachability-Based CBF}
\label{subsec:reachability cbf}

Constructive methods based on Hamilton-Jacobi (HJ) reachability enable the computation of value functions that characterize forward invariant sets~\cite{Lygeros2004}. In~\cite{Choi2021}, these methods are linked to the synthesis of value functions similar to CBFs, providing finite-horizon invariance guarantees. In the first part of this paper, we show how symmetries in the dynamics and constraints convey to symmetries in CBFs derived from HJ reachability (while broadening the scope in the second part). Specifically, we consider our reachability-based CBF definition~\cite{Wiltz2025b}, which (i) allows pointwise computation without requiring evaluation over an entire domain, (ii) guarantees forward invariance over infinite time horizons, and (iii) extends to certain time-varying constraints. It assumes the existence of a forward control invariant set and at least knowledge on one of its subsets. 

\begin{figure}[t]
	\centering
	\def\svgwidth{0.65\columnwidth}
\begingroup%
  \makeatletter%
  \providecommand\color[2][]{%
    \errmessage{(Inkscape) Color is used for the text in Inkscape, but the package 'color.sty' is not loaded}%
    \renewcommand\color[2][]{}%
  }%
  \providecommand\transparent[1]{%
    \errmessage{(Inkscape) Transparency is used (non-zero) for the text in Inkscape, but the package 'transparent.sty' is not loaded}%
    \renewcommand\transparent[1]{}%
  }%
  \providecommand\rotatebox[2]{#2}%
  \newcommand*\fsize{\dimexpr\f@size pt\relax}%
  \newcommand*\lineheight[1]{\fontsize{\fsize}{#1\fsize}\selectfont}%
  \ifx\svgwidth\undefined%
    \setlength{\unitlength}{413.59952671bp}%
    \ifx\svgscale\undefined%
      \relax%
    \else%
      \setlength{\unitlength}{\unitlength * \real{\svgscale}}%
    \fi%
  \else%
    \setlength{\unitlength}{\svgwidth}%
  \fi%
  \global\let\svgwidth\undefined%
  \global\let\svgscale\undefined%
  \makeatother%
  \begin{picture}(1,0.40710382)%
    \lineheight{1}%
    \setlength\tabcolsep{0pt}%
    \put(0,0){\includegraphics[width=\unitlength,page=1]{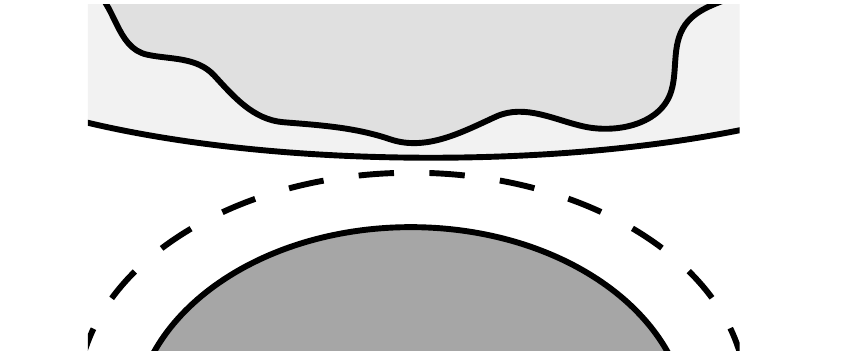}}%
    \put(0.38903693,0.04392397){\color[rgb]{0,0,0}\makebox(0,0)[lt]{\lineheight{1.25}\smash{\begin{tabular}[t]{l}{\small$h(x)<0$}\end{tabular}}}}%
    \put(0.38474308,0.15796462){\color[rgb]{0,0,0}\makebox(0,0)[lt]{\lineheight{1.25}\smash{\begin{tabular}[t]{l}{\small$h(x)>0$}\end{tabular}}}}%
    \put(0,0){\includegraphics[width=\unitlength,page=2]{setF_new.pdf}}%
    \put(0.05876518,0.17413546){\color[rgb]{0,0,0}\makebox(0,0)[lt]{\lineheight{1.25}\smash{\begin{tabular}[t]{l}$h(x)=\delta$\end{tabular}}}}%
    \put(0.46402124,0.33398824){\color[rgb]{0,0,0}\makebox(0,0)[lt]{\lineheight{1.25}\smash{\begin{tabular}[t]{l}$\mathcal{F}$\end{tabular}}}}%
    \put(0.15742607,0.2813495){\color[rgb]{0,0,0}\makebox(0,0)[lt]{\lineheight{1.25}\smash{\begin{tabular}[t]{l}$\mathcal{V}$\end{tabular}}}}%
  \end{picture}%
\endgroup%

	\caption{Illustration of Assumption~\ref{ass:setF},~\cite{Wiltz2025b}.}
	\label{fig:setF}
	\vspace{-\baselineskip}
\end{figure}

\begin{assumption}
	\label{ass:setF}
	There exists a forward control invariant subset $ \calV\subset\calH $ with respect to~\eqref{eq:dynamics} such that $ h(x)\geq \delta $ for all $ x\in\calV $ and some $ \delta > 0 $, see Figure~\ref{fig:setF}. While $ \calV $ is not required to be explicitly known, we assume that a closed subset $ \calF\subseteq\calV $ is known.
\end{assumption}

The next assumption ensures the existence of a finite time
\begin{subequations}
	\label{eq:T}
	\begin{align}
		T &\coloneq \sup_{x\in\calD\setminus\calF} \min_{\tau\geq 0} \tau \\
		\text{s.t.} \;\; & \dot{\bm{x}}(t) = f(\bm{x}(t),\bm{u}(t)) \quad \text{(a.e.)}, \\
		& \bm{x}(0) = x_{0}, \quad \bm{u}(t)\in\calU, \quad \bm{x}(\tau)\in\calF,
	\end{align}
\end{subequations}
that allows to determine the viability of any state in $ \calD $ by analyzing the system's behavior over a finite time horizon.

\begin{assumption}
	\label{ass:controllability}
	Let either of the following statements hold:
	\begin{enumerate}[leftmargin=0.9cm]
		\item[A\ref{ass:controllability}.1] Dynamics~\eqref{eq:dynamics} are controllable on the closure of $\calF^{\text{c}}$ where $\calF^{\text{c}}$ denotes the complement of $ \calF $; or
		\item[A\ref{ass:controllability}.2] For all $ x_{0}\in\calD\setminus\calF $, there exist $ t\geq 0 $ such that $ \calR_{t}(x_{0})\cap\calF \neq \emptyset $. 
	\end{enumerate}
\end{assumption}

With this, the reachability-based CBF becomes, for some $ \gamma>0 $ and the previously defined time-horizon $ T $,
\begin{subequations}
	\label{eq:finite horizon construction H}
	\begin{align}
		\label{seq:finite horizon construction H max min}
		H_{T}(x_{0}) &\coloneq \max_{\bm{u}(\cdot)\in\bm{\calU}_{[0,T]}} \min_{t\in[0,T]} h(\bm{x}(t)) - \gamma t \\
		\label{seq:finite horizon construction H initial condition}
		\text{s.t.}\;\; &\bm{x}(0)=x_{0},\\
		\label{seq:finite horizon construction H dynamics}
		&\dot{\bm{x}}(s) = f(\bm{x}(s),\bm{u}(s)) \quad (a.e.), \\
		\label{seq:finite horizon construction H input constraint}
		&\bm{u}(s)\in\calU, \qquad \forall s\in[0,T]\\
		\label{seq:finite horizon construction H terminal constraint}
		& \bm{x}(\vartheta)\in\calF, \qquad \text{for some } \vartheta\in[0,T].
	\end{align}
\end{subequations}

\begin{theorem}[\cite{Wiltz2025b}, Thm.~6]
	\label{thm:predictive CBF}
	Let Assumptions~\ref{ass:setF} and~\ref{ass:controllability} hold, and let $ h $ be Lipschitz continuous. Define $ T $ as in~\eqref{eq:T}, and let $ H_{T}: \calD\rightarrow\bbR $ be given by~\eqref{eq:finite horizon construction H} on a domain $ \calD\subseteq\bbR^{n} $ with $ \calH\subset\calD $. Assume $ \gamma \in [0, \delta/T] $, and that $ f $ is uniformly bounded on $ \calC $, i.e., for all $ x\in\calC $ there exists a $ u\in\calU $ such that $ ||f(x,u)||\leq M $ for some $ M>0 $. If $ H_{T} $ is locally Lipschitz continuous, then it is a CBF in the Dini sense on the domain~$ \calD $ with respect to dynamics~\eqref{eq:dynamics}. 
\end{theorem}


\section{Leveraging Symmetries and Equivariances in the CBF Synthesis}
\label{sec:equivarainces for symmetric constraints}

A major problem in the computation of value functions such as CBFs and CLFs is the curse of dimensionality, which is further aggravated for systems operating on large state spaces. In order to mitigate this problem, the \emph{symmetry properties of constraint set $ \calH $} and the \emph{equivariance properties of the dynamics~\eqref{eq:dynamics}} can be leveraged. With symmetries and equivariances, we are particularly referring to the following.

\begin{definition}[Symmetries in the constraints]
	\label{def:global symmetry}
	Recall $ \calH \coloneq \{x \, | \, h(x)\geq0\} $. Let $ D_{h}(\cdot;p): \bbR^{n} \rightarrow \bbR^{n} $ be some non-trivial isomorphism\footnote{An isomorphism $ D:\bbR^{n}\rightarrow\bbR^{n} $ is considered to be \emph{trivial} if it maps each point to itself, i.e., $ D(x) \equiv x $. It is \emph{non-trivial} if it is not trivial.} dependent on an (optional) parameter $ p\in\calP_{h}\subseteq\bbR^{n_{h}} $. We call $ h $ \emph{symmetric with respect to $ D_{h} $ and $ \calP_{h} $} if 
	\begin{align}
		\label{eq:symmetry 1}
		h(x) = h(D_{h}(x;p)) \qquad \forall p\in\calP_{h}.
	\end{align}
	Analogously, $ \calH $ is symmetric with respect to $ D_{h} $ and~$ \calP_{h} $ if
	\begin{align}
		\label{eq:symmetry 2}
		\calH = D_{h}(\calH;p) \qquad \forall p\in\calP_{h}.
	\end{align}
\end{definition}

\begin{definition}[Equivariant dynamics]
	\label{def:global invariance}
	Let $ D_{f}(\cdot;p): \bbR^{n}\rightarrow\bbR^{n} $ be a non-trivial diffeomorphism dependent on an (optional) parameter $ p\in\calP_{f}\subseteq\bbR^{n_{f}} $. We call the dynamics~\eqref{eq:dynamics} \emph{equivariant with respect to $ D_{f} $ and $ \calP_{f} $} if there exists some (possibly trivial) isomorphism $ D_{u}(\cdot;p)\!: \calU \rightarrow \calU $, with $ D_{u}(\calU;p)\subseteq\calU $ for all $ p\in\calP_{f} $ such that
	\begin{align}
		\label{eq:invariances 1}
		f(D_{f}(x;p),D_{u}(u;p)) = J_{D_{f}}\!(x;p) \; f(x,u) \qquad \!\forall 
		p\in\calP_{f},
	\end{align}
	where $ J_{D_{f}} $ denotes the Jacobian to $ D_{f} $ with respect to its first argument. We call~\eqref{eq:dynamics} \emph{strongly equivariant} if $ D_{u}(\calU;p)=\calU $ for all $ p\in\calP_{f} $.
\end{definition}

Intuitively, symmetry implies that if function $ h $ is known in some states $ x $, then $ h $ can be derived for further states via diffeomorphism $ D_{h}(\cdot;p) $ by varying $ p $ over parameter space~$ \calP_{h} $. The equivariance of dynamics~\eqref{eq:dynamics} intuitively means that the system's behavior is the same in any state $ x' = D_{f}(x;p) $, $ p\in\calP_{f} $. Examples for systems with equivariant dynamics are oscillators like a mechanical pendulum, or mobile systems such as unicycles or bicycles (see Section~\ref{sec:case studies}). 
Next, we establish how symmetries and equivariances convey to the symmetry properties of the reachability-based CBF $ H_{T} $ (Section~\ref{sec:beyond the symmetric case} generalizes the results to CBFs agnostic to the particular CBF synthesis method). Thereafter, we show how these properties can be exploited in the computation of~$ H_{T} $. 

\subsection{Symmetry Properties of the Reachability-Based CBF~$ \text{H}_{\text{T}} $}

CBFs are far from being unique even for the same system and state constraint. This freedom can be leveraged in their design and computation. Of particular interest in this respect are CBFs that inherit the symmetries of system dynamics and state constraints. One such function is the CBF~$ H_{T} $ from Section~\ref{subsec:reachability cbf}, which is a representative of the group of reachability-based and predictive CBFs~\cite{Wiltz2025b}. In particular, if the constraint specification $ h $ is symmetric and $ f $ equivariant with respect to the same diffeomorphism $ D $, then the CBF $ H_{T} $ defined in~\eqref{eq:finite horizon construction H} possesses favorable symmetry properties. 

\begin{theorem}[Symmetry of $ H_{T} $]
	\label{thm:symmetry H_T}
	Let Assumptions~\ref{ass:setF} and~\ref{ass:controllability} hold, and let $ H_{T}: \calD\rightarrow\bbR $ be defined in~\eqref{eq:finite horizon construction H} with $ h $ and $ T $ being the same as in Theorem~\ref{thm:predictive CBF}. Moreover, let there be a non-trivial diffeomorphism $ D(\cdot;p): \bbR^{n}\rightarrow\bbR^{n} $ with an (optional) parameter $ p\in\calP $ such that
	\begin{enumerate}[label=(\arabic*)]
		\item $ h $ is \emph{symmetric} with respect to $ D $ and $ \calP $;
		\item dynamics~\eqref{eq:dynamics} are \emph{strongly equivariant} with respect to $ D $ and $ \calP $;
		\item there exists a set $ \calF $ satisfying Assumption~\ref{ass:setF} such that $ D(\calF;p) = \calF $ for any $ p\in\calP $
		(i.e., $ \calF $ has the same symmetry properties as $ \calH $, cf. \eqref{eq:symmetry 2}).
	\end{enumerate} 
	Then, $ H_{T}(x) = H_{T}(D(x;p)) $ for any $ p\in\calP $ and all $ x\in\calD $.
\end{theorem}
\begin{proof}
	Let us define $ \tilde{x}_{0} \coloneq D({x}_{0};p) $ for some $ x_{0}\in\calD $ and $ p\in\calP $. With $ \tilde{x}_{0} $ as initial state, optimization problem \eqref{eq:finite horizon construction H} becomes
	\begin{subequations}
		\label{eq:thm:symmetry H_T aux 1}
		\begin{align}
			\label{seq:thm:symmetry H_T aux 1.1}
			H_{T}(D(x_{0};&p)) \coloneq \max_{\bm{u}(\cdot)\in\bm{\calU}_{[0,T]}} \min_{t\in[0,T]} h(\tilde{\bm{x}}(t)) - \gamma t \\
			\label{seq:thm:symmetry H_T aux 1.2}
			\text{s.t.}\;\; &\tilde{\bm{x}}(0)=\tilde{x}_{0} \coloneq D({x}_{0};p),\\
			\label{seq:thm:symmetry H_T aux 1.3}
			&\dot{\tilde{\bm{x}}}(s) = f(\tilde{\bm{x}}(s),\tilde{\bm{u}}(s)) \quad \text{(a.e.)}, \\
			\label{seq:thm:symmetry H_T aux 1.4}
			&\tilde{\bm{u}}(s)\in\calU \qquad \forall s\in[0,T],\\
			\label{seq:thm:symmetry H_T aux 1.5}
			& \tilde{\bm{x}}(\vartheta)\in\calF \qquad \text{for some } \vartheta\in[0,T].
		\end{align}
	\end{subequations}
	We now proceed as follows: By employing $ \tilde{x} = D(x;p) $, $ x = D^{-1}(\tilde{x};p) $ and $ \tilde{u} = D_{u}(u;p) $, where $ D_{u} $ is an isomorphism as by Definition~\ref{def:global invariance}, as well as the symmetry and equivariance properties of  $ h $, $ \calF $ and $ f $, respectively, we derive for the optimization objective and each of the constraints in~\eqref{eq:thm:symmetry H_T aux 1} equivalent expressions. From these, we conclude the equivalence of \eqref{eq:thm:symmetry H_T aux 1} and~\eqref{eq:finite horizon construction H}, which establishes the result. 
	
	\emph{Optimization objective~\eqref{seq:thm:symmetry H_T aux 1.1}:} Starting with the argument of the max-min-problem on the left-hand side, we derive
	\begin{align}
		\label{eq:thm:symmetry H_T aux 2}
		h(\tilde{\bm{x}}(t)) - \gamma t = h(D(\bm{x}(t);p)) - \gamma t \stackrel{\eqref{eq:symmetry 1}}{=} h(\bm{x}(t)) - \gamma t,
	\end{align}
	where the latter equality holds due to the symmetry of $ h $. 
	
	\emph{Constraint~\eqref{seq:thm:symmetry H_T aux 1.2}:} By applying $ D^{-1}(\cdot;p) $ to both sides of~\eqref{seq:thm:symmetry H_T aux 1.2}, we obtain an expression equivalent to~\eqref{seq:finite horizon construction H initial condition} as
	\begin{align}
		\label{eq:thm:symmetry H_T aux 3}
		\bm{x}(0) = D^{-1}(\tilde{\bm{x}}(0);p) \stackrel{\eqref{seq:thm:symmetry H_T aux 1.2}}{=} D^{-1}( D({x}_{0};p);p) = x_{0}.
	\end{align}
	\emph{Constraint~\eqref{seq:thm:symmetry H_T aux 1.3}:} By using $ \tilde{x}_{0} \coloneq D({x}_{0};p) $, we derive for the left- and right-hand sides of~\eqref{seq:thm:symmetry H_T aux 1.3}, respectively, 
	\begin{align*}
		\dot{\tilde{\bm{x}}}(s) &= \frac{dD}{ds}(\bm{x}(s);p) = J_{D}(\bm{x}(s);p) \; \dot{\bm{x}}(s)\\
		f(\tilde{\bm{x}}(s),\tilde{\bm{u}}(s)) &= f(D(\bm{x}(s);p),D_{u}(\bm{u}(s);p)) \\
		&\stackrel{\eqref{eq:invariances 1}}{=} J_{D}(\bm{x}(s);p) \;  f(\bm{x}(s);\bm{u}(s))
	\end{align*}
	where the last equality holds due to the equivariance of~\eqref{eq:dynamics}. By comparing the right-hand sides of the two equations, which are by~\eqref{seq:thm:symmetry H_T aux 1.3} equal to each other, we conclude that
	\begin{align}
		\label{eq:thm:symmetry H_T aux 4}
		\dot{\bm{x}}(s) = f(\bm{x}(s),\bm{u}(s))
	\end{align}
	is equivalent to~\eqref{seq:thm:symmetry H_T aux 1.3}. 
	
	\emph{Constraint~\eqref{seq:thm:symmetry H_T aux 1.4}:} We derive in a straightforward manner 
	\begin{align}
		\bm{u}= D_{u}^{-1}(\tilde{\bm{u}}(s);p) \stackrel{\eqref{seq:thm:symmetry H_T aux 1.4}}{\in} D_{u}^{-1}(\calU;p) = \calU,
	\end{align}
	where the latter equality follows from the strong equivariance of dynamics~\eqref{eq:dynamics}.
	
	\emph{Constraint~\eqref{seq:thm:symmetry H_T aux 1.5}:}  By also applying $ D^{-1}(\cdot;p) $ to both sides of~\eqref{seq:thm:symmetry H_T aux 1.5}, we obtain an expression equivalent to~\eqref{seq:thm:symmetry H_T aux 1.5} as 
	\begin{align}
		\label{eq:thm:symmetry H_T aux 5}
		\bm{x}(\vartheta) = D^{-1}(\tilde{\bm{x}}(\vartheta);p) \stackrel{\eqref{seq:thm:symmetry H_T aux 1.5}}{\in} D^{-1}(\calF;p) = \calF,
	\end{align}
	where the last equality holds as $ \calF $ is assumed to have the same symmetry properties as $ \calH $.
	
	Summarizing~\eqref{eq:thm:symmetry H_T aux 2}-\eqref{eq:thm:symmetry H_T aux 5}, we obtain from \eqref{eq:thm:symmetry H_T aux 1} the equivalent optimization problem
	\begin{align*}
		H_{T}(D(x_{0};p)) &\coloneq \max_{\bm{u}(\cdot)\in\bm{\calU}_{[0,T]}} \min_{t\in[0,T]} h(\bm{x}(t)) - \gamma t \\
		\text{s.t.}\;\; \bm{x}(0)&=x_{0},\; \\
		\bm{x}(x)&=f(\bm{x}(s), \bm{u}(s)) \quad \text{(a.e.)},\\ 
		\bm{u}(s)&\in\calU \qquad \forall s\in[0,T],\; \\ 
		\bm{x}(\theta) &\in \calF  \qquad \text{for some } \vartheta\in[0,T],
	\end{align*}
	which in turn is equivalent to~\eqref{eq:finite horizon construction H}. Thus, $ H_{T}(x) = H_{T}(D(x;p)) $ for any $ p\in\calP $ and all $ x\in\calD $.
\end{proof}

\begin{figure}[t]
	\centering
	\def\svgwidth{0.5\columnwidth}
\begingroup%
  \makeatletter%
  \providecommand\color[2][]{%
    \errmessage{(Inkscape) Color is used for the text in Inkscape, but the package 'color.sty' is not loaded}%
    \renewcommand\color[2][]{}%
  }%
  \providecommand\transparent[1]{%
    \errmessage{(Inkscape) Transparency is used (non-zero) for the text in Inkscape, but the package 'transparent.sty' is not loaded}%
    \renewcommand\transparent[1]{}%
  }%
  \providecommand\rotatebox[2]{#2}%
  \newcommand*\fsize{\dimexpr\f@size pt\relax}%
  \newcommand*\lineheight[1]{\fontsize{\fsize}{#1\fsize}\selectfont}%
  \ifx\svgwidth\undefined%
    \setlength{\unitlength}{113.75288475bp}%
    \ifx\svgscale\undefined%
      \relax%
    \else%
      \setlength{\unitlength}{\unitlength * \real{\svgscale}}%
    \fi%
  \else%
    \setlength{\unitlength}{\svgwidth}%
  \fi%
  \global\let\svgwidth\undefined%
  \global\let\svgscale\undefined%
  \makeatother%
  \begin{picture}(1,0.4983866)%
    \lineheight{1}%
    \setlength\tabcolsep{0pt}%
    \put(0,0){\includegraphics[width=\unitlength,page=1]{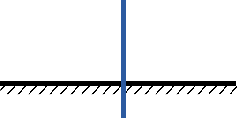}}%
    \put(0.53961071,0.43166268){\color[rgb]{0,0,0}\makebox(0,0)[lt]{\lineheight{1.25}\smash{\begin{tabular}[t]{l}{$\calM$}\end{tabular}}}}%
    \put(0,0){\includegraphics[width=\unitlength,page=2]{transformation_on_calM.pdf}}%
    \put(0.14287373,0.36880179){\color[rgb]{0,0,0}\makebox(0,0)[lt]{\lineheight{1.25}\smash{\begin{tabular}[t]{l}$D(x;p(x))$\end{tabular}}}}%
    \put(0.69986844,0.17001447){\color[rgb]{0,0,0}\makebox(0,0)[lt]{\lineheight{1.25}\smash{\begin{tabular}[t]{l}\footnotesize{$h(x)>0$}\end{tabular}}}}%
    \put(0.69947717,0.03877201){\color[rgb]{0,0,0}\makebox(0,0)[lt]{\lineheight{1.25}\smash{\begin{tabular}[t]{l}\footnotesize{$h(x)<0$}\end{tabular}}}}%
    \put(0.0766329,0.21197002){\color[rgb]{0,0,0}\makebox(0,0)[lt]{\lineheight{1.25}\smash{\begin{tabular}[t]{l}$x$\end{tabular}}}}%
    \put(0.56278151,0.2227901){\color[rgb]{0,0,0}\makebox(0,0)[lt]{\lineheight{1.25}\smash{\begin{tabular}[t]{l}$\tilde{x}$\end{tabular}}}}%
  \end{picture}%
\endgroup%

	\caption{Transformation of a point $ x $ to set $ \calM $ for an exemplary constraint $ h $ defined as the distance to the indicated obstacle. Constraint function $ h $ is symmetric with respect to a lateral shift denoted by~$ D $.}
	\label{fig:transfromation_to_calM}
	\vspace{-\baselineskip}
\end{figure}

The symmetry properties of $ H_{T} $ can be now leveraged to simplify the computation of the CBF. To this end, let us consider some $ \calM\subset\bbR^{n} $ (potentially a manifold) and a differentiable function $ p: \calD \rightarrow \calP $ mapping any state in the domain to the parameter space such that 
\begin{align}
	\label{eq:exploiting symmetry}
	D(x;p(x)) \in \calM \qquad \forall x\in\calD\smallsetminus\calM,
\end{align}
where $ D $ is the diffeomorphism in Theorem~\ref{thm:symmetry H_T}. Thereby, the parameterized transformation $ D(x;p(x)) $ maps any state $ x\in\calD $ to~$ \calM $. Then, the function $ H_{T}: \calD \rightarrow \bbR $ can be defined on its entire domain $ \calD $ in terms of the values of $ H_{T} $ on $ \calM $ as 
\begin{align}
	\label{eq:symmetry based H_T definition}
	H_{T}(x) = \begin{cases}
		H_{T}(x) & \text{if } x\in\calM, \\
		H_{T}(D(x;p(x))) & \text{if } x\in\calD\smallsetminus\calM.
	\end{cases}
\end{align}
Such a symmetry-based definition of $ H_{T} $ has the advantage that the numeric values of $ H_{T} $ only need to be computed on $ \calM $. All further values of $ H_{T} $ in any state $ x\in\calD $ are then given via diffeomorphism~$ D $ as by~\eqref{eq:symmetry based H_T definition}. The concept is illustrated for an exemplary constraint in Figure~\ref{fig:transfromation_to_calM}, where the values of $ H_{T} $ only need to be known on the set marked in blue.

\subsection{Local Symmetry Properties}

As symmetry and equivariance properties may not hold as globally as assumed so far, we shall briefly introduce a local variant of the previous result. Therefore, let us consider a local notion of the symmetry and equivariance properties. 

\begin{definition}[Local symmetry of $ h $ and $ \calH $]
	\label{def:local symmetry}
	Let $ D_{h} $ and $ \calP_{h} $ be the same as in Definition~\ref{def:global symmetry}, and let $ \calL\subseteq\bbR^{n} $. We call $ h $ \emph{locally symmetric on $ \calL $ with respect to $ D_{h} $ and $ \calP_{h} $} if 
	\begin{align}
		\label{eq:local symmetry 1}
		h(x) = h(D_{h}(x;p)) \qquad \forall x\in\calL\cap D_{h}^{-1}(\calL;p) \quad \forall p\in\calP_{h}.
	\end{align}	
\end{definition}

\begin{figure}
	\centering
	\begin{minipage}{0.49\columnwidth}
		\centering
		\def\svgwidth{1\columnwidth}
\begingroup%
  \makeatletter%
  \providecommand\color[2][]{%
    \errmessage{(Inkscape) Color is used for the text in Inkscape, but the package 'color.sty' is not loaded}%
    \renewcommand\color[2][]{}%
  }%
  \providecommand\transparent[1]{%
    \errmessage{(Inkscape) Transparency is used (non-zero) for the text in Inkscape, but the package 'transparent.sty' is not loaded}%
    \renewcommand\transparent[1]{}%
  }%
  \providecommand\rotatebox[2]{#2}%
  \newcommand*\fsize{\dimexpr\f@size pt\relax}%
  \newcommand*\lineheight[1]{\fontsize{\fsize}{#1\fsize}\selectfont}%
  \ifx\svgwidth\undefined%
    \setlength{\unitlength}{112.34636442bp}%
    \ifx\svgscale\undefined%
      \relax%
    \else%
      \setlength{\unitlength}{\unitlength * \real{\svgscale}}%
    \fi%
  \else%
    \setlength{\unitlength}{\svgwidth}%
  \fi%
  \global\let\svgwidth\undefined%
  \global\let\svgscale\undefined%
  \makeatother%
  \begin{picture}(1,0.56994439)%
    \lineheight{1}%
    \setlength\tabcolsep{0pt}%
    \put(0,0){\includegraphics[width=\unitlength,page=1]{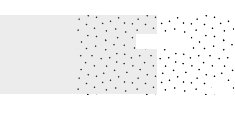}}%
    \put(0.91102513,0.19122854){\color[rgb]{0,0,0}\makebox(0,0)[lt]{\lineheight{1.25}\smash{\begin{tabular}[t]{l}\footnotesize{$\calL$}\end{tabular}}}}%
    \put(0.54594887,0.05066957){\color[rgb]{0,0,0}\makebox(0,0)[lt]{\lineheight{1.25}\smash{\begin{tabular}[t]{l}\footnotesize{$h(x)<0$}\end{tabular}}}}%
    \put(0.57980416,0.37674348){\color[rgb]{0,0,0}\makebox(0,0)[lt]{\lineheight{1.25}\smash{\begin{tabular}[t]{l}\footnotesize{$D_{h}(\cdot;p)$}\end{tabular}}}}%
    \put(0.10798726,0.56286788){\color[rgb]{0,0,0}\makebox(0,0)[lt]{\lineheight{1.25}\smash{\begin{tabular}[t]{l}\footnotesize{$h(D_{h}(x;p))=h(x)$ holds}\end{tabular}}}}%
    \put(0.00024793,0.18141029){\color[rgb]{0,0,0}\makebox(0,0)[lt]{\lineheight{1.25}\smash{\begin{tabular}[t]{l}\footnotesize{$D_{h}^{-1}(\calL;p)$}\end{tabular}}}}%
    \put(0,0){\includegraphics[width=\unitlength,page=2]{local_symmetry.pdf}}%
  \end{picture}%
\endgroup%

		\caption{Local symmetry with $ \calL $ being the dotted and $ D_{h}^{-1}(\calL;p) $ the gray region.}
		\label{fig:local_symmetry}
		\vspace{-\baselineskip}
	\end{minipage}%
	\hfill
	\begin{minipage}{0.45\columnwidth}
		\centering
		\def\svgwidth{0.75\columnwidth}
		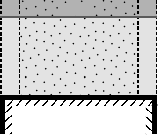
		\caption{Local symmetry of~$ H_{T} $.}
		\label{fig:local_cbf_symmetry}
		\vspace{-\baselineskip}
	\end{minipage}
\end{figure}

\begin{remark}
	The (global) symmetry of $ h $ in the sense of Definition~\ref{def:global symmetry} is equivalent to $ h $ being locally symmetric on $ \calL = \bbR^{n} $. The term $ D_{h}^{-1}(\calL;p) $ in~\eqref{eq:local symmetry 1} restricts the definition to those points $ x $ that are mapped to $ \calL $ by $ D_{h} $, where the symmetry property holds. The set $ \calL\cap D_{h}^{-1}(\calL;p) $ is denoted by the gray-shaded dotted region in Figure~\ref{fig:local_symmetry}.
\end{remark}

An illustration of the concept of local symmetry can be found in Figure~\ref{fig:local_symmetry}. In this example, $ h $ is defined as the distance to the indicated obstacle, which is locally symmetric on $ \calL $ with respect to a lateral shift denoted by $ D_{h} $. Similarly, we define a local notion of equivariance.

\begin{definition}[Local equivariance] 
	\label{def:local invariance}
	We call the dynamics~\eqref{eq:dynamics} \emph{locally equivariant on $ \calL\subseteq\bbR^{n} $ with respect to $ D_{f} $ and $ \calP_{f} $}, where $ D_{f} $, $ \calP_{f} $ are the same as in Definition~\ref{def:global invariance}, if there exists some (possibly trivial) isomorphism $ D_{u}(\cdot;p)\!: \calU \rightarrow \calU $, with $ D_{u}(\calU;p)\subseteq\calU $ for all $ p\in\calP $, such that
	\begin{align}
		f(D_{f}(x;p),D_{u}(u;p)) = J_{D_{f}}\!(x;p) \, f(x,u) \quad \forall x\!\in\!\calL \quad \forall 
		p\!\in\!\calP_{f}.
	\end{align}
	We call~\eqref{eq:dynamics} \emph{locally strongly equivariant} if $ D_{u}(\calU;p)=\calU $ for all~$ p\in\calP $.
\end{definition}

\begin{remark}
	The local notion of equivariance defined here is equivalent to that in Definition~\ref{def:global invariance} for $ \calL = \bbR^{n} $.
\end{remark}

A local version of Theorem~\ref{thm:symmetry H_T} can be obtained if the symmetry and equivariance properties in the theorem hold at least locally on a subset $ \calL\subseteq\calD $ of the domain, and the state trajectory $ \bm{\varphi}(\cdot;x_{0},\bm{u}_{0}^{\ast}) $ stays within $ \calL $ for all $ x_{0} $ contained in some subset $ \calL_{0}\subseteq\calL $; see Figure~\ref{fig:local_cbf_symmetry} for an illustration. Then, the symmetry property of $ H_{T} $ can be still obtained on $ \calL_{0} $, which corresponds to the dotted area in Figure~\ref{fig:local_cbf_symmetry}. 

\begin{theorem}[Local symmetry of $ H_{T} $]
	\label{thm:local symmetry H_{T}}
	Let Assumptions~\ref{ass:setF} and~\ref{ass:controllability} hold, and let $ H_{T}:\calD \rightarrow \bbR $ be defined in~\eqref{eq:finite horizon construction H} with $ h $ and $ T $ being the same as in Theorem~\ref{thm:predictive CBF}. Moreover, let $ \calL $ and $ \calL_{0} $ be sets with $ \calL_{0} \subseteq \calL \subseteq \calD $ such that for all $ x_{0}\in\calL_{0} $
	\begin{align}
		\label{eq:thm:local symmetry H_{T}}
		\bm{\varphi}(t;x_{0},\bm{u}_{x_{0}}^{\ast}) \in\calL \qquad \forall t\in[0,T]
	\end{align}
	where $ \bm{u}_{x_{0}}^{\ast} $ denotes the optimal input trajectory solving~\eqref{eq:finite horizon construction H}. If there exists a non-trivial diffeomorphism $ D(\cdot;p): \bbR^{n} \rightarrow \bbR^{n} $ with an (optional) parameter $ p\in\calP $ such that 
	\begin{enumerate}[label=(\arabic*)]
		\item $ h $ is \emph{locally symmetric} on $ \calL $ with respect to diffeomorphism $ D $ and parameters $ \calP $,
		\item dynamics $ f $ are \emph{locally strongly equivariant} on $ \calL $ with respect to diffeomorphism $ D $ and parameters $ \calP $,
		\item there exists a set $ \calF $ satisfying Assumption~\ref{ass:setF} such that $ D(\calF;p)\cap\calL = \calF\cap\calL $ for any $ p\in\calP $ (i.e., $ \calF $ has the same symmetry properties on $ \calL $ as $ \calH $),
	\end{enumerate}
	then it holds $ H_{T}(x) = H_{T}(D(x;p)) $ for all $ x\in\calL_{0} $ and $ p\in\calP $.
\end{theorem}
\begin{proof}
	This result is obtained by using arguments analogous to those in the proof of Theorem~\ref{thm:symmetry H_T}.
\end{proof}

This result is of interest for the computation of $ H_{T} $ as well since it allows for a simplified computation of the CBF on $ \calL_{0} $ similarly to the global case. In particular, $ H_{T} $ can be expressed anywhere on $ \calL_{0} $ in terms of its values on some subset $ \calM\subset\calL_{0} $ if there exists a differentiable function $ p:\calL_{0}\rightarrow\calP $ such that 
\begin{align*}
	D(x;p(x)) \in\calM \qquad \forall x\in\calL_{0}\smallsetminus\calM.
\end{align*}
Thereby, we obtain
\begin{align}
	\label{eq:local symmetry based H_T definition}
	H_{T}(x) = \begin{cases}
		H_{T}(x) & \text{if } x\in\calM, \\
		H_{T}(D(x;p(x))) & \text{if } x\in\calL_{0}\smallsetminus\calM.
	\end{cases}
\end{align}

To motivate~\eqref{eq:thm:local symmetry H_{T}}, let us briefly consider the following example. 

\begin{example}
Let a vehicle be given with turning radius $ R $ and a strictly positive forward speed. Given a set $ \calL $, then it is straightforward that the trajectory $ \bm{\varphi}(\cdot;x_{0},\bm{u}_{x_{0}}^{\ast}) $ stays within $ \calL $ for any initial state $ x_{0}\in\calL_{0} \coloneq \calL \ominus \calB_{2\!R} $ and \eqref{eq:thm:local symmetry H_{T}} is satisfied.
\end{example}

\section{Case Studies on Symmetries and Equivariances}
\label{sec:case studies}

\begin{figure}
	\centering
	\begin{subfigure}{0.49\columnwidth}
		\centering
		\def\svgwidth{0.71\columnwidth}
\begingroup%
  \makeatletter%
  \providecommand\color[2][]{%
    \errmessage{(Inkscape) Color is used for the text in Inkscape, but the package 'color.sty' is not loaded}%
    \renewcommand\color[2][]{}%
  }%
  \providecommand\transparent[1]{%
    \errmessage{(Inkscape) Transparency is used (non-zero) for the text in Inkscape, but the package 'transparent.sty' is not loaded}%
    \renewcommand\transparent[1]{}%
  }%
  \providecommand\rotatebox[2]{#2}%
  \newcommand*\fsize{\dimexpr\f@size pt\relax}%
  \newcommand*\lineheight[1]{\fontsize{\fsize}{#1\fsize}\selectfont}%
  \ifx\svgwidth\undefined%
    \setlength{\unitlength}{265.07514053bp}%
    \ifx\svgscale\undefined%
      \relax%
    \else%
      \setlength{\unitlength}{\unitlength * \real{\svgscale}}%
    \fi%
  \else%
    \setlength{\unitlength}{\svgwidth}%
  \fi%
  \global\let\svgwidth\undefined%
  \global\let\svgscale\undefined%
  \makeatother%
  \begin{picture}(1,0.90116535)%
    \lineheight{1}%
    \setlength\tabcolsep{0pt}%
    \put(0,0){\includegraphics[width=\unitlength,page=1]{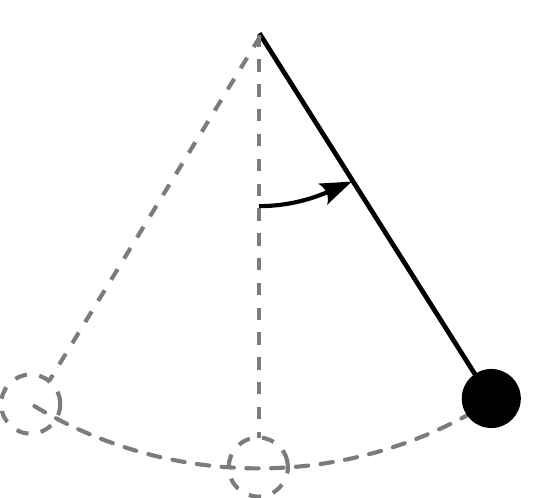}}%
    \put(0.7586832,0.40285433){\color[rgb]{0,0,0}\makebox(0,0)[lt]{\lineheight{1.25}\smash{\begin{tabular}[t]{l}\footnotesize{$l$}\end{tabular}}}}%
    \put(0.53717196,0.8126891){\color[rgb]{0,0,0}\makebox(0,0)[lt]{\lineheight{1.25}\smash{\begin{tabular}[t]{l}\footnotesize{$d_m$}\end{tabular}}}}%
    \put(0.50394525,0.57316366){\color[rgb]{0,0,0}\makebox(0,0)[lt]{\lineheight{1.25}\smash{\begin{tabular}[t]{l}\footnotesize{$\theta$}\end{tabular}}}}%
    \put(0,0){\includegraphics[width=\unitlength,page=2]{pendulum_sketch.pdf}}%
    \put(0.93770926,0.07150916){\color[rgb]{0,0,0}\makebox(0,0)[lt]{\lineheight{1.25}\smash{\begin{tabular}[t]{l}\footnotesize{$g$}\end{tabular}}}}%
    \put(0,0){\includegraphics[width=\unitlength,page=3]{pendulum_sketch.pdf}}%
  \end{picture}%
\endgroup%

		\caption{}
		\label{fig:pendulum_sketch}
	\end{subfigure}%
	\hfill
	\begin{subfigure}{0.49\columnwidth}
		\centering
		\def\svgwidth{0.9\columnwidth}
\begingroup%
  \makeatletter%
  \providecommand\color[2][]{%
    \errmessage{(Inkscape) Color is used for the text in Inkscape, but the package 'color.sty' is not loaded}%
    \renewcommand\color[2][]{}%
  }%
  \providecommand\transparent[1]{%
    \errmessage{(Inkscape) Transparency is used (non-zero) for the text in Inkscape, but the package 'transparent.sty' is not loaded}%
    \renewcommand\transparent[1]{}%
  }%
  \providecommand\rotatebox[2]{#2}%
  \newcommand*\fsize{\dimexpr\f@size pt\relax}%
  \newcommand*\lineheight[1]{\fontsize{\fsize}{#1\fsize}\selectfont}%
  \ifx\svgwidth\undefined%
    \setlength{\unitlength}{196.33143628bp}%
    \ifx\svgscale\undefined%
      \relax%
    \else%
      \setlength{\unitlength}{\unitlength * \real{\svgscale}}%
    \fi%
  \else%
    \setlength{\unitlength}{\svgwidth}%
  \fi%
  \global\let\svgwidth\undefined%
  \global\let\svgscale\undefined%
  \makeatother%
  \begin{picture}(1,0.77607508)%
    \lineheight{1}%
    \setlength\tabcolsep{0pt}%
    \put(0,0){\includegraphics[width=\unitlength,page=1]{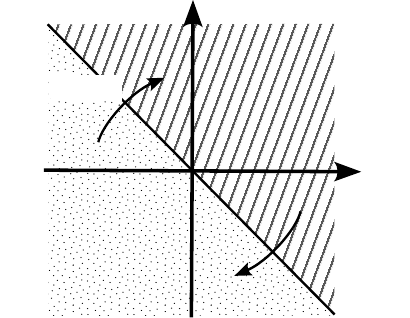}}%
    \put(0.12813208,0.53871512){\color[rgb]{0,0,0}\makebox(0,0)[lt]{\lineheight{1.25}\smash{\begin{tabular}[t]{l}\footnotesize{$D(x)$}\end{tabular}}}}%
    \put(0,0){\includegraphics[width=\unitlength,page=2]{pendulum_symmetry.pdf}}%
    \put(0.49227287,0.47458452){\color[rgb]{0,0,0}\makebox(0,0)[lt]{\lineheight{1.25}\smash{\begin{tabular}[t]{l}\footnotesize{$x_2\geq -x_1$}\end{tabular}}}}%
    \put(0.56890283,0.55960431){\color[rgb]{0,0,0}\makebox(0,0)[lt]{\lineheight{1.25}\smash{\begin{tabular}[t]{l}$\calM$\end{tabular}}}}%
    \put(0.82041243,0.28623009){\color[rgb]{0,0,0}\makebox(0,0)[lt]{\lineheight{1.25}\smash{\begin{tabular}[t]{l}\footnotesize{$x_1$}\end{tabular}}}}%
    \put(0.49711149,0.73019293){\color[rgb]{0,0,0}\makebox(0,0)[lt]{\lineheight{1.25}\smash{\begin{tabular}[t]{l}\footnotesize{$x_2$}\end{tabular}}}}%
    \put(-0.01530269,0.73669045){\color[rgb]{0,0,0}\makebox(0,0)[lt]{\lineheight{1.25}\smash{\begin{tabular}[t]{l}\footnotesize{$x_2= -x_1$}\end{tabular}}}}%
  \end{picture}%
\endgroup%

		\caption{}
		\label{fig:pendulum_M}
	\end{subfigure}
	\caption{Mechanical pendulum (disturbed): (a)~sketch, and (b)~set $ \calM $.}
	\label{fig:pendulum}
	\vspace{-\baselineskip}
\end{figure}

Before generalizing our findings, we investigate symmetries and equivariances for exemplary constraints and systems to illustrate the application of our results so far. 

\subsection{Mechanical Pendulum}
Consider a mechanical pendulum as depicted in Figure~\ref{fig:pendulum_sketch} subject to an input disturbance in terms of an unknown, possibly time-varying angular momentum $ d_{m} $, where $ |d_{m}|\leq d_{\text{max}} $ with some positive constant $ d_{\text{max}} $~\cite{Wiltz2024a}. The dynamics of the pendulum are given as 
\begin{align}
	\label{eq:pendulum}
	\begin{bmatrix}
		\dot{x}_{1} \\ \dot{x}_{2}
	\end{bmatrix} &= f_{1}(x,u) \coloneq \begin{bmatrix}
		x_{2} \\
		-\frac{g}{l} \, \sin(x_{1}) + d_{m} + u
	\end{bmatrix}.
\end{align}
Here, $ x_{1} = \theta $ and $ x_{2}=\dot{\theta} $ denote the angle of excitation and its velocity, respectively, and $ l>0 $ the length of the pendulum. The system is strongly equivariant with respect to $ D(x) = \begin{bsmallmatrix} -x_{1} \\ -x_{2} \end{bsmallmatrix} $ as 
\begin{align*}
	f_{1}(D(x),D_{u}(u)) = \begin{bsmallmatrix}
		-x_{2} \\
		\! +\frac{g}{l} \sin(x_{1}) + d_{m} + D_{u}(u) \! 
	\end{bsmallmatrix} = J_{D}(x) \, f_{1}(x,u),
\end{align*}
where we choose $ D_{u}(u) = -u $, and where we exploit the symmetry of the set of admissible disturbances $ d_{m} $. We point out that the additional transformation $ D_{u} $ allows to establish the equality between left- and right-hand side in the latter equation. The requirement $ D_{u}(\calU)=\calU $ is satisfied for any input constraint of the form $ \calU \coloneq \{u \, | \, |u|\leq u_{\text{max}}\} $, $ u_{\text{max}}>0. $  

Furthermore, consider a constraint function $ h $ with $ h(x) = h(-x) $, for example that of a rotated ellipse
\begin{align}
	\label{eq:h for pendulum}
	 h(x) = -\sqrt{\tfrac{(x_{1}+x_{2})^{2}}{2a^{2}} + \tfrac{(-x_{1}+x_{2})^{2}}{2b^{2}}} + 1
\end{align}
for some constants $ a, b\in\bbR_{>0} $. Such constraint is symmetric with respect to $ D $. Then, we can leverage the symmetry property of the constraint and the equivariance of the dynamics to determine $ H_{T} $ as a CBF. In particular, when choosing $ \calM = \{ x \, | \, x_{2} \geq -x_{1} \} $, then $ H_{T} $ can be expressed through $ D $ in terms of its values on $ \calM $ as detailed in~\eqref{eq:symmetry based H_T definition}; an illustration of $ \calM $ is given in Figure~\ref{fig:pendulum_M}. The resulting CBF, computed for the exemplary constraint~\eqref{eq:h for pendulum} with $ a=1 $, $ b=2 $, is depicted in Figure~\ref{fig:cbf_mechanical_pendulum}. As it can be seen, the resulting CBF is symmetric under a rotation of 180~degrees, which corresponds to the symmetry induced by transformation~$ D $.

\subsection{Kinematic Bicycle Model}
\label{subsec:exampel Kinematic bicycle model}

\begin{figure}[t]
	\begin{minipage}{0.55\columnwidth}
		\centering
		\includegraphics[width=\linewidth]{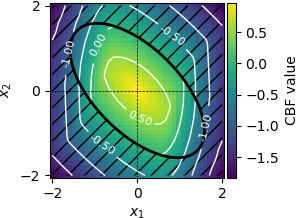}
		\caption{CBF for the mechanical pendulum~\eqref{eq:pendulum} and the corresponding state constraint.}
		\label{fig:cbf_mechanical_pendulum}
		\vspace{-\baselineskip}
	\end{minipage}
	\hfill
	\begin{minipage}{0.4\columnwidth}
		\centering
		\def\svgwidth{1.0\columnwidth}
\begingroup%
  \makeatletter%
  \providecommand\color[2][]{%
    \errmessage{(Inkscape) Color is used for the text in Inkscape, but the package 'color.sty' is not loaded}%
    \renewcommand\color[2][]{}%
  }%
  \providecommand\transparent[1]{%
    \errmessage{(Inkscape) Transparency is used (non-zero) for the text in Inkscape, but the package 'transparent.sty' is not loaded}%
    \renewcommand\transparent[1]{}%
  }%
  \providecommand\rotatebox[2]{#2}%
  \newcommand*\fsize{\dimexpr\f@size pt\relax}%
  \newcommand*\lineheight[1]{\fontsize{\fsize}{#1\fsize}\selectfont}%
  \ifx\svgwidth\undefined%
    \setlength{\unitlength}{256.26495556bp}%
    \ifx\svgscale\undefined%
      \relax%
    \else%
      \setlength{\unitlength}{\unitlength * \real{\svgscale}}%
    \fi%
  \else%
    \setlength{\unitlength}{\svgwidth}%
  \fi%
  \global\let\svgwidth\undefined%
  \global\let\svgscale\undefined%
  \makeatother%
  \begin{picture}(1,0.81626357)%
    \lineheight{1}%
    \setlength\tabcolsep{0pt}%
    \put(0,0){\includegraphics[width=\unitlength,page=1]{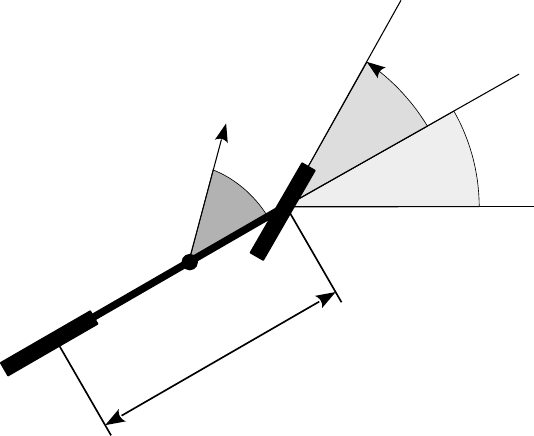}}%
    \put(0.39607649,0.40454786){\color[rgb]{0,0,0}\makebox(0,0)[lt]{\lineheight{1.25}\smash{\begin{tabular}[t]{l}$\beta$\end{tabular}}}}%
    \put(0.64354382,0.56033633){\color[rgb]{0,0,0}\makebox(0,0)[lt]{\lineheight{1.25}\smash{\begin{tabular}[t]{l}$\zeta$\end{tabular}}}}%
    \put(0.78059458,0.4530023){\color[rgb]{0,0,0}\makebox(0,0)[lt]{\lineheight{1.25}\smash{\begin{tabular}[t]{l}$\psi$\end{tabular}}}}%
    \put(0.43136955,0.06876114){\color[rgb]{0,0,0}\makebox(0,0)[lt]{\lineheight{1.25}\smash{\begin{tabular}[t]{l}$L$\end{tabular}}}}%
    \put(0.34575181,0.234386){\color[rgb]{0,0,0}\makebox(0,0)[lt]{\lineheight{1.25}\smash{\begin{tabular}[t]{l}$C$\end{tabular}}}}%
    \put(0.32587123,0.43125159){\color[rgb]{0,0,0}\makebox(0,0)[lt]{\lineheight{1.25}\smash{\begin{tabular}[t]{l}$v$\end{tabular}}}}%
    \put(0,0){\includegraphics[width=\unitlength,page=2]{bicycle.pdf}}%
  \end{picture}%
\endgroup%

		\caption{Kinematic bicycle model.}
		\label{fig:kinematic bicycle model}
		\vspace{-\baselineskip}
	\end{minipage}
\end{figure}

Let us now consider the kinematic bicycle model~\cite{Wang2001} (see Figure~\ref{fig:kinematic bicycle model}) given as
\begin{subequations}
	\label{eq:bicycle model}
	\begin{align}
		\label{seq:bicycle xdot}
		\dot{x} &= v \cos(\psi + \beta(\zeta)) \\
		\label{seq:bicycle ydot}
		\dot{y} &= v \sin(\psi + \beta(\zeta)) \\
		\label{seq:bicycle psidot}
		\dot{\psi} &= \frac{v \cos(\beta(\zeta))\tan(\zeta)}{L} 
	\end{align}
\end{subequations}
where $ \beta(\zeta) = \arctan(\frac{1}{2}\tan(\zeta)) $. The position of the center of mass $ C $ is denoted by $ \mathbf{x}_{\text{pos}}=[x, y]^{T} $, and the vehicle's orientation by $ \psi $; inputs are velocity~$ v $ and steering angle~$ \zeta $. The stack vector of the system's states is denoted by $ \mathbf{x} = [x,y,\psi]^{T} $. The kinematic bicycle model allows for various useful equivariances.

\subsubsection{Translational Equivariance}
\label{paragraph:bicycle translational invariance}
At first, we observe that the system is strongly equivariant with respect to translations in its positional states $ x $, $ y $ given as 
\begin{align*}
	D_{1}(\mathbf{x};p) \coloneq \begin{bsmallmatrix}
		x + p_{x} \\ y + p_{y} \\ \psi
	\end{bsmallmatrix},
\end{align*}
where $ p = \begin{bsmallmatrix}
	p_{x} \\ p_{y}
\end{bsmallmatrix} \in\bbR^{2} $. This is the case as the right-hand side of~\eqref{eq:bicycle model} is independent of the positional states. 

\begin{figure}
	\centering
	\def\svgwidth{0.39\columnwidth}
	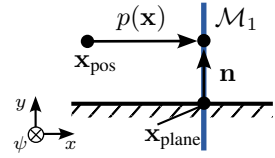
	\caption{Translational equivariance and symmetry: illustration of diffeomorphism~$ D_{1} $ and transformation $ p $.}
	\label{fig:bicycle_trafo_translation}
	\vspace{-\baselineskip}
\end{figure}

Now, let constraint set $ \calH $ be an \emph{arbitrary half-plane}. Such~$ \calH $ is symmetric with respect to $ D_{1} $ and parameter set $ \calP_{1} \coloneq \{ p \, | \, \langle \mathbf{n}, p\rangle = 0\} $. To see this, let us write the half-plane in terms of constraint function $ h(\mathbf{x}) =\langle\mathbf{n},\mathbf{x}_{\text{pos}} - \mathbf{x}_{\text{plane}}\rangle $, where $ \mathbf{n}\in\bbR^{2} $ denotes the unit normal vector to the plane, and $ \mathbf{x}_{\text{plane}}\in\bbR^{2} $ is some arbitrary point on the $ x $-$ y $-plane. Then, $ h $ is symmetric with respect to $ D_{1} $ and $ \calP_{1} $ as $ h(D_{1}(\mathbf{x};p)) = h(\mathbf{x}) $ for any $ p \in \calP_{1} $. Thereby, Theorem~\ref{thm:symmetry H_T} is applicable. Hence, when the numerical values of $ H_{T} $ are known on 
\begin{align*}
	\calM_{1} = \{ \mathbf{x}\in\calD \, | \, \mathbf{x}_{\text{pos}} = \mathbf{x}_{\text{plane}} + \nu \mathbf{n}, \; \nu\in\bbR, \; \psi \in [0,2\pi) \},
\end{align*}
then $ H_{T} $ is given on its entire domain $ \calD $ by~\eqref{eq:symmetry based H_T definition} for 
\begin{align*}
	p(\mathbf{x}) = - \langle R(-\pi/2)\mathbf{n}, \mathbf{x}_{\text{pos}} - \mathbf{x}_{\text{plane}} \rangle R(-\pi/2)\mathbf{n},
\end{align*}
where the rotation matrix $ R $ is given as $ R(\rho) \coloneq \begin{bsmallmatrix}
	\cos(\rho) & -\sin(\rho) \\ \sin(\rho) & \cos(\rho)
\end{bsmallmatrix} $. It is straightforward to verify that $ p(\mathbf{x})\in\calP_{1} $ and $ D_{1}(\mathbf{x};p(\mathbf{x})) = \left[\begin{smallmatrix}
\!\mathbf{x}_{\text{pos}} + p(\mathbf{x})\! \\ \psi
\end{smallmatrix}\right] \in\calM_{1} $ as required. The transformation of state $ \mathbf{x} $ into set $ \calM_{1} $ is illustrated in Figure~\ref{fig:bicycle_trafo_translation}.

\subsubsection{Rotational Equivariance} 
\begin{figure}
	\centering
	\begin{subfigure}{0.49\columnwidth}
		\centering
		\def\svgwidth{0.65\columnwidth}
		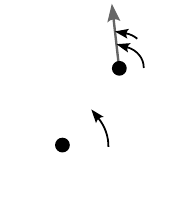
		\caption{}
		\label{fig:bicycle_cos_trafo_rotation}
	\end{subfigure}%
	\hfill
	\begin{subfigure}{0.49\columnwidth}
		\centering
		\def\svgwidth{0.75\columnwidth}
		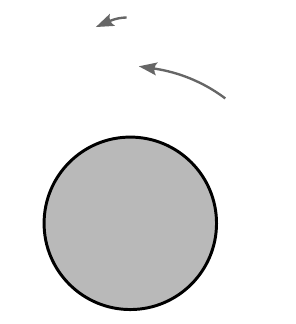
		\caption{}
		\label{fig:bicycle_application_trafo_rotation}
	\end{subfigure}
	\caption{Rotational equivariance and symmetries: (a)~illustration of the coordinate transformation, and (b)~application of diffeomorphism~$ D_{2} $ to a circular obstacle.}
	\label{fig:bicycle_rotation}
	\vspace{-\baselineskip}
\end{figure}

Next, we show that the kinematic bicycle model is equivariant with respect to a rotation around a fixed point as illustrated in Figure~\ref{fig:bicycle_rotation}. To establish the result formally, transform dynamics~\eqref{eq:bicycle model} to a polar coordinate-based system with states $ \chi = [r, \varphi, \theta]^{T} $ (see Figure~\ref{fig:bicycle_cos_trafo_rotation}) by using diffeomorphism $ \begin{bsmallmatrix}
	r \\ \varphi \\ \psi
\end{bsmallmatrix} \mapsto \begin{bsmallmatrix}
	x_{\text{c}} + r \cos(\varphi) \\ y_{\text{c}} + r \sin(\varphi) \\ \varphi + \theta
\end{bsmallmatrix}
 = 
\begin{bsmallmatrix}
	x \\ y \\ \psi
\end{bsmallmatrix} $ for $ x \neq x_{c} $, $ y\neq y_{c} $. Here, $ x_{c}, y_{c}\in\bbR $ are constants defining the center around which $ r $ and $ \varphi $ specify the bicycle's positions. With this, the transformed dynamics~\eqref{eq:bicycle model} are given as
\begin{align}
	\label{eq:bicycle model polar}
	\left[
	\begin{smallmatrix}
		\dot{r} \\ 
		\dot{\varphi} \\ 
		\dot{\theta} 
	\end{smallmatrix} \right] =
	\left[
	\begin{smallmatrix}
		v \cos(\theta + \beta(\zeta)) \\ 
		\frac{v}{r} \sin(\theta+\beta(\zeta)) \\ 
		\frac{1}{L} v \cos(\beta(\zeta)) \tan(\zeta) - \frac{1}{r} v\sin(\theta+\beta(\zeta))
	\end{smallmatrix}\right].
\end{align}
Since the right-hand side is independent of $ \varphi $, \eqref{eq:bicycle model polar} is strongly equivariant with respect to $ D_{2}(\chi;p) \coloneq [r, \, \varphi \!+\! p, \, \theta]^{T} $, $ p \in\bbR $. 

Now, consider a circular constraint set $ \calH $ centered in $ (x_{c}, y_{c}) $. Such $ \calH $ is symmetric with respect to $ D_{2} $ and $ \calP_{2} \coloneq [0,2\pi) $. To see this, consider $ h(\mathbf{x}) = ||\mathbf{x}_{\text{pos}} - \begin{bsmallmatrix}
	x_{\text{c}} \\ y_{\text{c}}
\end{bsmallmatrix}||^{2} - \bar{r}^{2} $, or equivalently, $ h(\mathbf{\chi}) = r^{2} - \bar{r}^{2} $ with $ \bar{r}\in\bbR_{>0} $ being the radius of the obstacle; the symmetry property follows immediately from $ h $ as it is independent of~$ \varphi $. Thereby, Theorem~\ref{thm:symmetry H_T} is applicable. Thus, when the numerical values of $ H_{T} $ are known on $ \calM_{2} \coloneq \{ \mathbf{\chi} \in \calD \, | \, \varphi = \varphi_{0} \} $ for some $ \varphi_{0} \in [0,2\pi) $, then $ H_{T} $ is given on $ \calD $ by~\eqref{eq:symmetry based H_T definition} for $ p(\mathbf{\chi}) = \varphi_{0} - \varphi $ (see Figure~\ref{fig:bicycle_application_trafo_rotation}). 

\subsubsection{Orientational Equivariance} 
\begin{figure}
	\centering%
	\begin{subfigure}{0.44\columnwidth}
		\centering
		\def\svgwidth{1\columnwidth}
		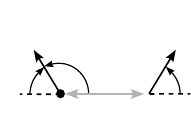
		\caption{}
		\label{fig:bicycle_orientational_invariance_trafo}
	\end{subfigure}
	\hfill
	\begin{subfigure}{0.49\columnwidth}
		\centering
		\def\svgwidth{1\columnwidth}
		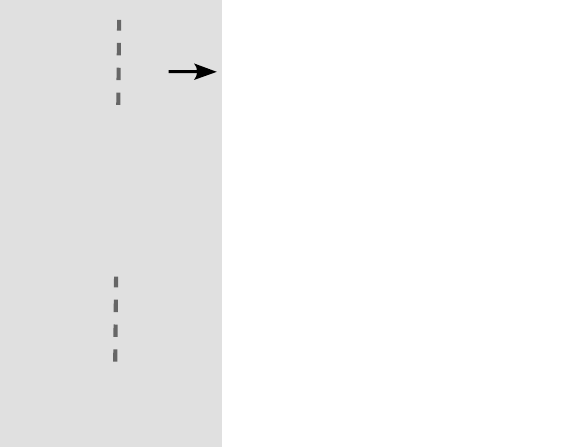
		\caption{}
		\label{fig:bicycle_orientational_invariance_derivation_diffeo}
	\end{subfigure}
	\caption{Equivariance and symmetry with respect to the orientation: (a)~illustration of the diffeomorphism~$ D_{3} $ for $ \rho = 0 $, and (b)~the derivation of diffeomorphism~$ D_{3} $.}
	\label{fig:bicycle_orientation}
	\vspace{-\baselineskip}
\end{figure}

The kinematic bicycle model also exhibits equivariances with respect to orientation. To this end, consider the diffeomorphism that mirrors a state around a vector $ \mathbf{n}\in\bbR^{2} $, normal to some constraint at a point $ \mathbf{x}_{\text{p}} $ (see Figure~\ref{fig:bicycle_orientational_invariance_trafo}). This transformation is useful for determining $ H_{T} $ in cases involving mirror-symmetric constraints, such as half-planes or circular boundaries. From a computational point of view, it allows to halve the effort for computing $ H_{T} $. 
To derive the respective diffeomorphism, write normal vector $ \mathbf{n} $ as $ \mathbf{n} = R(\rho) \mathbf{e}_{2} $ in terms of the standard unit vector $ \mathbf{e}_{2} \coloneq \begin{bsmallmatrix}
	0 \\ 1
\end{bsmallmatrix} $ and a unique angle $ \rho \in [0,2\pi) $; $ R $ denotes the rotation matrix as before. 
The diffeomorphism, denoted by $ D_{3} $, is constructed as the composition of three distinct diffeomorphisms $ D_{3}(\mathbf{x};\rho) = D_{III} \circ D_{II} \circ D_{I}(\mathbf{x};\rho) $ (see Figure~\ref{fig:bicycle_orientational_invariance_derivation_diffeo}), which yields overall
\begin{align*}
	D_{3}(\mathbf{x};\rho) = \begin{bsmallmatrix}
		[\bm{I} - 2\cos(\rho) R(\rho)] (\mathbf{x}_{\text{pos}}-\mathbf{x}_{\text{p}}) + \mathbf{x}_{\text{p}} \\
		\pi - \psi
	\end{bsmallmatrix}.
\end{align*}
By substituting $ D_{3} $ and $ D_{u}(\begin{bsmallmatrix}
	v \\ \zeta
\end{bsmallmatrix}) = \begin{bsmallmatrix}
	v \\ -\zeta
\end{bsmallmatrix} $ into~\eqref{eq:invariances 1}, the equivariance of the kinematic bicycle model~\eqref{eq:bicycle model} with respect to $ D_{3} $ and $ \rho\in\calP_{3}\coloneq [0,2\pi) $ follows. If the input constraint on the steering angle $ \zeta $ is furthermore of the form $ |\zeta| < \zeta_{\text{max}} $, then~\eqref{eq:bicycle model} is even strongly equivariant with respect to $ D_{3} $. 

Now, reconsider constraint function $ h(\mathbf{x})\!=\!\langle\mathbf{n},\mathbf{x}_{\text{pos}} \!-\! \mathbf{x}_{\text{plane}}\rangle $, which is also symmetric with respect to $ D_{3} $. To see this, observe that $ D_{3} $ can be equivalently written as $ D_{3}(\mathbf{x};\rho) = \begin{bsmallmatrix}
	\mathbf{x}_{\text{pos}} - 2 \, \langle \mathbf{e}_{1},R(-\rho) (\mathbf{x}_{\text{pos}}-\mathbf{x}_{\text{p}})\rangle \, R(\rho)\mathbf{e}_{1} \\ 
	\pi - \psi
\end{bsmallmatrix} $, and thus $ h(D_{3}(\mathbf{x};\rho)) \!=\! \langle \mathbf{n}, \mathbf{x}_{\text{pos}} \!-\! \mathbf{x}_{\text{plane}} \rangle \!-\! 2 \, \langle \mathbf{e}_{1},R(-\rho) (\mathbf{x}_{\text{pos}}\!-\!\mathbf{x}_{\text{p}})\rangle \, \langle \mathbf{n}, R(\rho)\mathbf{e}_{1}\rangle \!=\! h(\mathbf{x}), $ where $ \langle \mathbf{n}, R(\rho)\mathbf{e}_{1}\rangle \!=\! \langle R(\rho) \mathbf{e}_{2}, R(\rho) \mathbf{e}_{1} \rangle \!=\! 0 $. From this, symmetry follows. Thereby, Theorem~\ref{thm:symmetry H_T} is also here applicable. Consequently, if in Part~1) the numeric values of $ H_{T} $ are known on $ \calM'_{1} = \{ \mathbf{x}\in\calM_{1} \, | \, \psi\in[0,\pi)\} $, then the remaining values of $ H_{T} $ on $ \calM_{1} $ are given by $ H_{T}(\mathbf{x}) = H_{T}(D_{3}(\mathbf{x},\rho)) $, $ \mathbf{x}\in\calM_{1}\setminus\calM'_{1} $. Thereby, the effort for the computation of $ H_{T} $ in Part~1) can be further reduced by $ 50\% $ with this equivariance. 

\vspace{0.5\baselineskip}

The equivariances of the kinematic bicycle model derived here will be utilized in the numerical examples presented in Section~\ref{sec:numeric examples}. Similar equivariances can be also derived for other vehicles; the derivation is analogous.

\subsection{Linear Systems}
\label{subsec:linear system}

Consider a linear system given for $ A\!\in\!\bbR^{n\!\times\! n} $, $ B\!\in\!\bbR^{n\!\times\! m} $ as
\begin{align}
	\label{eq:linear dynamics}
	\dot{x} = Ax + Bu, \qquad x(0) = x_{0}.
\end{align}
Such system is equivariant with respect to a linear diffeomorphism if and only if the following holds.

\begin{proposition}
	\label{prop:linear invariance general}
	Consider diffeomorphism 
	\begin{align}
		\label{eq:linear diffeo}
		D(x;p,\Delta x) \coloneq D_{p} \, x + \Delta x,
	\end{align}
	where $ p\in\calP\subseteq\bbR^{n_{p}} $ and $ \Delta x\in\calP_{\Delta x}\subseteq\bbR^{n} $ are parameters, and $ D_{p}\in\bbR^{n\times n} $ is a matrix dependent on parameter $ p $ and invertible for all $ p\in\calP $. The linear dynamics~\eqref{eq:linear dynamics} are equivariant with respect to $ D $ and $ \calP_{D} \times \calP_{\Delta x} $ if and only if all of the following conditions hold: 
	\begin{enumerate}[label=(\arabic*)]
		\item $ A $ and $ D_{p} $ commute, i.e., $ A D_{p} = D_{p} A $;
		\item for all $ p\in\calP $, there exist $ D_{u,p} \in \bbR^{m\times m} $ such that $ D_{p} B = B D_{u,p} $ and $ D_{u,p} \,\calU \subseteq \calU $;
		\item $ \Delta x \in \calN(A) $, where $ \calN(A) $ denotes the null space of $ A $.
	\end{enumerate}
\end{proposition}
\begin{proof}
	The sufficiency of the conditions follows immediately by evaluating~\eqref{eq:invariances 1} and substituting the conditions, which yields
	\begin{align*}
		&A D(x;p,\Delta x) + B D_{u,p} u = A D_{p} x + A \Delta x + B D_{u,p} u \\
		&\qquad= D_{p} A x + D_{p} B u = D_{p} (Ax + Bu).
	\end{align*}
	Necessity follows by equating the coefficients. 
\end{proof}

In the particular case that $ D_{p} $ is diagonalizable, we obtain a class of constraints that is symmetric with respect to~\eqref{eq:linear diffeo}.

\begin{proposition}
	\label{prop:linear systems symmetric constraints}
	Let $ D $ be as defined in~\eqref{eq:linear diffeo}, and let $ C\in \bbR $ be some constant. Furthermore, let us denote the \mbox{$ i $-th} left-eigenvector of $ D_{p} $ by $ w_{i} $, the respective eigenvalue by $ \lambda_{i}(D_{p}) $, and let $ \Delta x \in \{v \, | \, w_i \!\perp\! v, \; \forall i\!: \lambda_i(D_{p}) \!=\! 1\} $. A function $ h: \bbR^{n} \rightarrow \bbR $ is symmetric with respect to~$ D $ if 
	\begin{align}
		\label{eq:prop:linear systems symmetric constraints 1}
		h(x) = \sum_{\substack{i:\\ \lambda_{i}(D_{p})=1}} c_{i} \, w_{i}^{T} x + C
	\end{align}
	with any constants $ c_{i}\in\bbR $.
\end{proposition}
\begin{proof}
	This directly follows by verifying~\eqref{eq:symmetry 1} as
	\begin{align*}
		h(D(x;p,\Delta x)) &= \sum_{\substack{i:\\\lambda_{i}(D_{p})=1}} c_{i} \, w_{i}^{T} (D_{p} \, x + \Delta x) + C \\
		&= \sum_{\substack{i:\\ \lambda_{i}(D_{p})=1}} c_{i} w_{i}^{T} x + C = h(x),
	\end{align*}
	where $ w_{i}^{T} \Delta x = 0 $ as $ \Delta x \perp w_{i} $ for any $ i $ with $ \lambda_{i}(D_{p}) = 1 $.
\end{proof}

\begin{figure}
 	\centering
 	\begin{subfigure}{0.46\columnwidth}
 		\centering
 		\def\svgwidth{1\columnwidth}
 		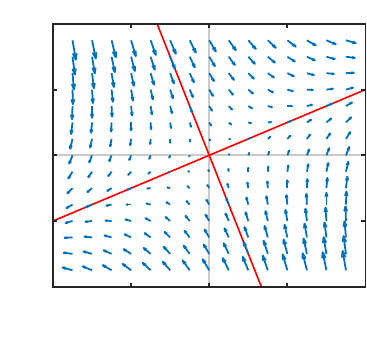
 		\caption{}
 		\label{fig:linear_system_slope_with_symmetries}
 	\end{subfigure}%
 	\hfill
 	\begin{subfigure}{0.49\columnwidth}
 		\centering
 		\includegraphics[width=\linewidth]{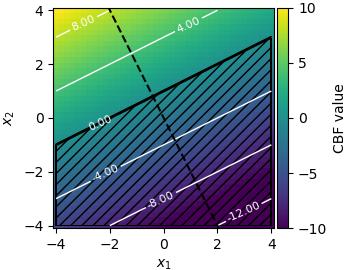}
 		\caption{}
 		\label{fig:fig:linear_system_H_T_heatmap}
 	\end{subfigure}
 	\caption{Linear system in Example~\ref{example:linear system}: (a)~its flow for $ u=0 $ with symmetries indicated, and (b)~the corresponding symmetric CBF~$ H_{T} $ for constraint $ h(x) \coloneq [-1 \;\, 2]\, x + 1\geq0 $ with symmetry axis indicated by the dashed line.}
 	\label{fig:linear_system_exmp}
 	\vspace{-\baselineskip}
\end{figure}

We apply Propositions~\ref{prop:linear invariance general} and~\ref{prop:linear systems symmetric constraints} in the following example.

\begin{example}
	\label{example:linear system}
	Consider diffeomorphism~\eqref{eq:linear diffeo} with $ D_{p} \coloneq P \, \text{diag}(p) \, P^{-1} $, $ P = \begin{bsmallmatrix}
		-1 & 2 \\ 3 & 1
	\end{bsmallmatrix} $, $ \calP \subseteq \bbR^{2} $ and $ \Delta x = 0 $, as well as linear system $ \dot{x} = \begin{bsmallmatrix}
		1 & 2 \\ 3 & -4
	\end{bsmallmatrix} x + \begin{bsmallmatrix}
	-1 \\ 3
	\end{bsmallmatrix} u $.	Based on Proposition~\ref{prop:linear invariance general}, we verify that the systems is equivariant with respect to $ D $. Noting that the left-eigenvalues to $ D_{p} $ are $ w_{1} = [-1,2] $ and $ w_{2} = [3,1] $, it follows by Proposition~\ref{prop:linear systems symmetric constraints}, that $ h(x) = w_{i}^{T} x + C $, for any $ i\in\{1,2\} $, is symmetric with respect to $ D $ and $ \calP = \{p \, | \, p = \begin{bsmallmatrix}
		1 \\ p_{2}
	\end{bsmallmatrix}, \; p_{2}\in\bbR \} $ if $ i=1 $, and $ \calP = \{p \, | \, p = \begin{bsmallmatrix}
		p_{1} \\ 1
	\end{bsmallmatrix}, \; p_{2}\in\bbR \} $ if $ i=2 $. As the system is now shown to be equivariant and the constraints symmetric, Theorem~\ref{thm:symmetry H_T} can be applied. Thereby, exemplarily for $ i=1 $, $ H_{T} $ is defined in terms of its values on $ \calM = \{ x \, | \, x = \sigma_{1} w_{1}^{T} + \sigma_{2} w_{2}^{T}, \; \sigma_{1}\in\bbR, \; \sigma_{2} \in \bbR_{\geq 0} \} $ through~\eqref{eq:symmetry based H_T definition} for $ p(x) = \begin{bsmallmatrix}
		1 \\ -1
	\end{bsmallmatrix} $. The resulting symmetry of $ H_{T} $ can be seen from Figure~\ref{fig:fig:linear_system_H_T_heatmap}.
\end{example} 

\begin{figure}
	\centering
	\begin{subfigure}{0.4\columnwidth}
		\centering
		\def\svgwidth{1\columnwidth}
		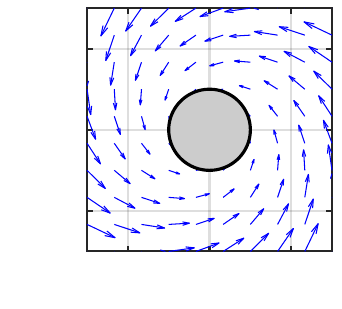
		\caption{}
		\label{fig:linear_system_rotational}
	\end{subfigure}%
	\hfill
	\begin{subfigure}{0.49\columnwidth}
		\centering
		\includegraphics[width=0.9\linewidth]{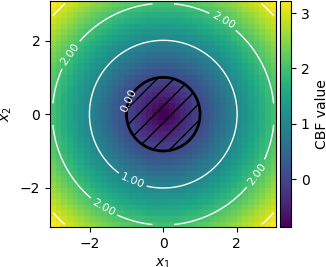}
		\caption{}
		\label{fig:linear_system rotational_H_T_heatmap}
	\end{subfigure}
	\caption{Exemplary linear system equivariant with respect to rotation $ R(\rho) $: (a)~its flow for $ u=0 $, and (b)~the corresponding symmetric CBF~$ H_{T} $ for constraint $ h(x) \coloneq ||x||-1\geq0 $.}
	\label{fig:linear_system_rotational_exmp}
	\vspace{-\baselineskip}
\end{figure}

Another class of linear systems are those that are equivariant with respect to rotations. For example, using Proposition~\ref{prop:linear invariance general}, it can be readily verified that~\eqref{eq:linear dynamics} is equivariant with respect to $ D(x;\rho) = R(\rho)x $, where $ R $ denotes the rotational matrix as previously defined, provided that $ A $ takes the form $ A = \begin{bsmallmatrix}
	a & b \\ -b & a
\end{bsmallmatrix} $, $ a,b\in\bbR $, and $ B $ is invertible. An example is given in Figure~\ref{fig:linear_system_rotational_exmp} for $ A = \begin{bsmallmatrix}
	-1 & -2 \\ 2 & -1
\end{bsmallmatrix} $, $ B=\bm{I}_{2} $, input constraint $ ||u||\leq3 $ and state constraint $ h(x) = ||x|| - 1 \geq 0 $.
Another such system is the double integrator; its dynamics in two dimensions are $ \dot{x} = \begin{bsmallmatrix}
	\bm{0}_{2} & \bm{I}_{2} \\ \bm{0}_{2} & \bm{0}_{2}
\end{bsmallmatrix} x + \begin{bsmallmatrix}
	\bm{0}_{2} \\ \bm{I}_{2}
\end{bsmallmatrix} u $. 
Consider diffeomorphism $ D(x;\rho) = \begin{bsmallmatrix}
	R(\rho) & \bm{0}_{2} \\ \bm{0}_{2} & R(\rho)
\end{bsmallmatrix} x $, and $ h(x) = x^{T} \begin{bsmallmatrix}
	\bm{I}_{2} & \bm{0}_{2} \\ \bm{0}_{2} & \bm{0}_{2}
\end{bsmallmatrix} x - r^{2} \geq 0 $, $ r\in\bbR $, as a circular constraint. By verifying the conditions in Proposition~\ref{prop:linear invariance general}, we conclude that the double integrator is equivariant, and $ h $ is symmetric with respect to $ D $ and $ \rho\in\bbR $ allowing the application of Theorem~\ref{thm:symmetry H_T}. 
We provide a numeric example on the double-integrator in Section~\ref{sec:numeric examples}.

\section{Leveraging Equivariances Beyond Symmetric Constraints}
\label{sec:beyond the symmetric case}

The results in the previous section relied on the fact that there exists a diffeomorphism $ D $ under which both the dynamics are equivariant \emph{and} the state constraint symmetric. Under these favorable premises, we were able to exactly synthesize CBF $ H_{T} $ on its entire domain based on its numerical values on a subset of its domain. On the other hand, such diffeomorphisms only exist for particular constraints, and the choice of constraints is thereby limited due to the dynamics. This section generalizes our previous approach to a broader class of constraints. Furthermore, the now developed results are independent of the particular CBF synthesis method. 

More specifically, we show in the course of this section that methods similar to those in the previous sections remain applicable as long as the dynamics are equivariant, even when the state constraints lack symmetry properties. Notably, we do no longer require the symmetry of $ h $ and the strong equivariance of the dynamics. This enhances the versatility of our approach in addressing a multitude of constraints. 

In the sequel, we assume an initial CBF~$ b $ to be given, which we leverage to synthesize new CBFs for state constraints beyond the one that $ b $ has been designed for. We do this by exploiting the equivariances of the system dynamics. This leads to an equivariance-based synthesis method similar to that introduced in Section~\ref{sec:equivarainces for symmetric constraints}.
We begin by assuming that $ b $ is fully known over its entire domain. Subsequently, we extend the results to accomodate cases, where only partial knowledge of $ b $ is available.

\subsection{Equivariance-Based Synthesis with Complete Knowledge of a Given CBF}

\begin{figure}[t]
	\centering
	\def\svgwidth{0.6\columnwidth}
	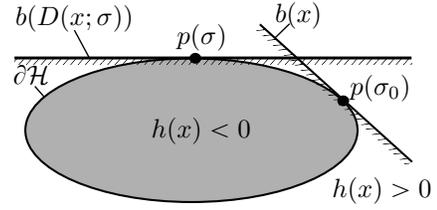
	\caption{Conceptual CBF synthesis method via equivariances for more general constraints.}
	\label{fig:approximation_idea}
	\vspace{-\baselineskip}
\end{figure}

Suppose a CBF $ b: \bbR^{n} \rightarrow \bbR $, as defined in Definition~\ref{def:cbf dini}, is given. The CBF may be selected as $ H_{T} $ in~\eqref{eq:finite horizon construction H}, as previously discussed. Nevertheless, the subsequent analysis is not restricted to this particular choice; any other CBF is equally admissible. Furthermore, let dynamics~\eqref{eq:dynamics} be equivariant with respect to some diffeomorphism $ D(\cdot;\sigma) $ with parameter $ \sigma\in\calP $; strong equivariance is explicitly not required. Then, 
\begin{align*}
	b_{\sigma}(\cdot) \coloneq b(D(\cdot;\sigma))
\end{align*}
is a CBF as well, which we establish in the next lemma.

\begin{lemma}
	\label{lem:b_sigma is a CBF}
	Let $ b:\bbR^{n}\rightarrow\bbR $ be some CBF in the Dini sense as defined in Definition~\ref{def:cbf dini}. Moreover, let dynamics~\eqref{eq:dynamics} be equivariant with respect to a diffeomorphism~$ D(\cdot;\sigma) $, $ \sigma\in\calP $, as per Definition~\ref{def:global invariance}. Then, $ b_{\sigma} $ is a CBF in the Dini sense and it holds
	\begin{align}
		\label{eq:lem:b_sigma is a CBF 1}
		\sup_{u\in\calU} \{db_{\sigma}(x;f(x,u))\} \geq -\alpha(b_{\sigma}(x)),
	\end{align}
	where $ \alpha $ is the same extended class~$ \calK_{e} $ function as in Definition~\ref{def:cbf dini}.
\end{lemma}
\begin{proof}
	Since $ b $ is a CBF, it holds $ db(x;f(x,u)) \geq -\alpha(b(x)) $ by Definition~\ref{def:cbf dini}. For $ b_{\sigma}(x) $, we derive
	\begin{align*}
		db_{\sigma}(x;f(x,u)) &= db(D(x;\sigma);\frac{\partial D}{\partial x}(x;\sigma) \, f(x,u)) \\
		&\stackrel{\eqref{eq:invariances 1}}{=} db(D(x;\sigma);f(D(x;\sigma),\tilde{u})) \\
		&\stackrel{\eqref{eq:def cbf dini}}{\geq} -\alpha(b(D(x;\sigma))) = -\alpha(b_{\sigma}(x))
	\end{align*}
	where we used the equivariance of dynamics~\eqref{eq:dynamics} in terms of~\eqref{eq:invariances 1}, and where $ \tilde{u}\!=\!D_{u}(u;\sigma) $ for some isomorphism~$ D_{u} $. As $ \tilde{u}\!\in\! D_{u}(\calU;\sigma)\!\subseteq\!\calU $ for any $ \sigma\in\calP $ due to the equivariance of~\eqref{eq:dynamics} (see Definition~\ref{def:global invariance}), it follows that $ \sup_{u\in\calU}\{ db_{\sigma}(x;f(x,u)) \} \geq -\alpha(b_{\sigma}(x)) $. Thus, $ b_{\sigma} $ is a CBF. 
\end{proof}

Equipped with this, CBFs for a broad class of constraints can be constructed even if the constraints are not symmetric.
To this end, we write the boundary of constraint set $ \calH $ as a parameterized curve denoted by $ p: \bbR \rightarrow \partial\calH $; in the sequel, we denote the scalar argument of $ p $ by $ \sigma $. If such curve $ p $ allows for a diffeomorphism $ D $, under which the dynamics are equivariant, and that maps any $ p(\sigma) $ back to the starting point $ p(\sigma_{0}) $, then we can ``shift'' the known CBF along the boundary of the constraint set $ \calH $ as illustrated in Figure~\ref{fig:approximation_idea}. From this, a CBF is obtained.

\begin{theorem}
	\label{thm:outer approximation max cbf}
	Let $ b: \bbR^{n}\rightarrow\bbR $ be a given CBF in the Dini sense. Moreover, let dynamics~\eqref{eq:dynamics} be equivariant with respect to a diffeomorphism $ D(\cdot;\sigma): \bbR^{n} \rightarrow \bbR^{n} $ dependent on a scalar parameter $ \sigma\in\calP\subseteq\bbR $. Then, 
	\begin{align}
		\label{eq:CBF max definition}
		B(x) \coloneq \max_{\sigma\in\calP} b_{\sigma}(x)
	\end{align}  
	is a CBF in the Dini sense on $ \bbR^{n} $ with respect to~\eqref{eq:dynamics} and its zero super-level set is
	\begin{align*}
		\calC = \bigcup_{\sigma\in\calP} \{ x\in\bbR^{n} \,|\, b(D(x;\sigma)) \geq 0 \}.
	\end{align*}
\end{theorem}
\begin{proof}
	Let us denote the local Lipschitz constant of $ b $ by $ L $, and that of $ D(\cdot;\sigma) $ with respect to $ x $ for any $ \sigma\in\calP $ by $ L_{D} $. Then, we directly derive for $ x'\in\calB_{\varepsilon}(x) $, $ \varepsilon>0 $, that
	\begin{align*}
		&|B(x)-B(x')| = \left|\max_{\sigma} b_{\sigma}(x) - \max_{\sigma} b_{\sigma}(x')\right| \\
		&\quad \leq \left|\max_{\sigma}(b_{\sigma}(x) - b_{\sigma}(x'))\right| \\
		&\quad \leq \max_{\sigma}\left|b(D(x;\sigma)) - b(D(x';\sigma))\right| \\
		&\quad \leq \max_{\sigma}\left( {L} \,|D(x;\sigma)-D(x';\sigma)| \right) \leq {L} \, L_{D} \, |x-x'|.
	\end{align*}
	Thus, $ B $ is locally Lipschitz continuous with Lipschitz constant~$ {L} L_{D} $. 
	
	Next, we show that~\eqref{eq:def cbf dini} holds for any $ x\in\bbR^{n} $. To this end, let us define for an arbitrary state $ x_{0}\in\bbR^{n} $
	\begin{align}
		\label{eq:thm:outer approximation max cbf aux 1}
		\sigma_{0} \coloneq \argmax_{\sigma\in\calP} b(D(x_{0};\sigma))
	\end{align}
	as the parameter for which $ B(x_{0}) = b_{\sigma_{0}}(x_{0}) $. If the right-hand side of~\eqref{eq:thm:outer approximation max cbf aux 1} yields a set, we assign to $ \sigma_{0} $ an arbitrary element of this set. Moreover, we note that by Lemma~\ref{lem:b_sigma is a CBF}, $ b_{\sigma} $ is a CBF and it holds $ db_{\sigma}(x;f(x,u)) \geq -\alpha(b_{\sigma}(x)) $ for some $ u\in\calU $. As by~\eqref{eq:CBF max definition}, it follows for some $ \varepsilon>0 $ that
	\begin{align*}
		B(x_{0}+\Delta x) \geq b_{\sigma_{0}}(x_{0}+\Delta x) \qquad \forall \Delta x \in \calB_{\varepsilon}(0)
	\end{align*}
	and we conclude that for some $ u\in\calU $ it holds
	\begin{align*}
		\begin{split}
			dB(x_{0};f(x_{0},u)) &\geq db_{\sigma_{0}}(x_{0};f(x_{0},u)) \\
			&\geq -\alpha(b_{\sigma_{0}}(x_{0})) = -\alpha(B(x_{0}))
		\end{split}
	\end{align*}
	Thereby, \eqref{eq:def cbf dini} is satisfied. 
	At last, we derive $ \calC $ as
	\begin{align*}
		\calC &= \{x \, | \, B(x)\geq 0\} = \{x \, | \, \max_{\sigma} b_{\sigma}(x) \geq 0\} \\
		&= \{x \, | \, \exists \sigma: b_{\sigma}(x) \geq 0\}  \\
		&= \bigcup_{\sigma} \{x \, | \, b(D(x;\sigma)) \geq 0\},
	\end{align*}
	which concludes the proof.
\end{proof}

\subsection{Equivariance-Based Synthesis under Partial Knowledge of a Given CBF}

The previous theorem requires full knowledge of CBF~$ b $. Therefore, the direct application of the previous theorem in the general case is challenging. However, a related result can be established under partial knowledge of $ b $. Specifically, only the following knowledge about $ b $ is required. 

\begin{assumption}[Partial Knowledge of a Given CBF]
	\label{ass:partial knowledge of a given CBF}
	Let $ b: \bbR^{n} \rightarrow \bbR $ be some CBF on a domain $ \calD\subseteq\bbR^{n} $, and let $ \calM\subseteq\calD $ be a set such that its $ \varepsilon $-neighborhood is entirely contained within $ \calD $; that is, there exists a constant $ \varepsilon_{\!\calM}>0 $ such that $ \calM\oplus\calB_{\varepsilon_{\!\calM}} \subseteq \calD $. The CBF $ b $ is assumed to be known and locally Lipschitz continuous on $ \calM\oplus\calB_{\varepsilon_{\!\calM}} $.
\end{assumption}

The pointwise computation method proposed in our earlier work~\cite{Wiltz2025b} is well suited to compute the numeric values of $ b $ on $ \calM\oplus\calB_{\varepsilon_{\!\calM}} $. However, also any other CBF synthesis method may be employed. Based on set~$ \calM $ and the previously defined diffeomorphism~$ D(\cdot;\sigma) $, $ \sigma\in\calP $, under which the dynamics~\eqref{eq:dynamics} are assumed to be equivariant, let us define the following sets: the transformed set $ \calM_{\sigma} \coloneq D^{-1}(\calM;\sigma) $, the union $ \widehat{\calM} \coloneq \bigcup_{\sigma\in\calP} \calM_{\sigma} $, and the neighborhood $ \calM_{\sigma}^{\varepsilon} \coloneq \bigcup_{|\Delta\sigma|<\varepsilon} \calM_{\sigma\!+\!\Delta\sigma} $. Without loss of generality, we assume that $ D $ is such that $ \calM = D(\calM;0) $.

\begin{figure}[t]
	\centering
	\def\svgwidth{0.6\columnwidth}
	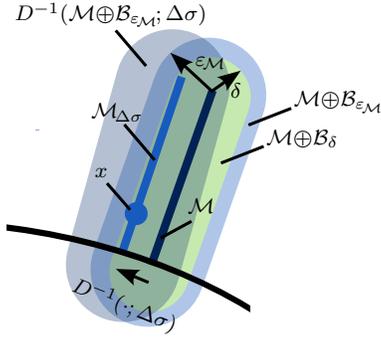
	\caption{Illustration of conditions~\eqref{eq:local minimum condition b under D}--\eqref{eq:vareps_M condition} for $ \sigma=0 $: In~\eqref{eq:local minimum condition b under D}, the values of $ b $ and $ b_{\Delta\sigma} $ are compared on the dark blue line representing set $ \calM $. Condition~\eqref{eq:Delta sigma condition 2} specifies the bounds for~$ \Delta\sigma $: for any point in $ \calM\oplus\calB_{\delta} $ (green), there must exist a $ \Delta\sigma $ within the $ \varepsilon $-bounds such that $ x\in\calM_{\Delta\sigma} $ (light-blue). Condition~\eqref{eq:vareps_M condition} ensures that the values of both $ b $ and $ b_{\Delta\sigma} $ are known on $ \calM $, where~\eqref{eq:local minimum condition b under D} compares their values.
	}
	\label{fig:Delta sigma condition}
	\vspace{-\baselineskip}
\end{figure}

To account for the partial knowledge of $ b $, we specify how its values under $ D $, namely the values of $ b_{\sigma}(\cdot) = b(D(\cdot;\sigma)) $, may vary as $ \sigma $ changes. In particular, we impose for all $ \sigma\in\calP $ the condition that
\begin{align}
	\label{eq:local minimum condition b under D}
	b(D(x;\sigma)) \geq b(D(x;\sigma\!+\!\Delta\sigma)) \!\qquad\! \forall x\!\in\!\calM_{\sigma}, \forall |\Delta\sigma|\!<\!\varepsilon
\end{align}
(condition on a local maximum in $ \sigma $ at fixed $ x\in\calM_{\sigma} $), where $ \varepsilon\coloneq\varepsilon(\sigma)>0 $ is chosen such that 
\begin{align}
	\label{eq:Delta sigma condition 2}
	x\in\calM_{\sigma}\!\oplus\!\calB_{\delta} \quad \Rightarrow \quad \exists |\Delta\sigma|\!<\!\varepsilon\!: \; x\!\in\!\calM_{\sigma\!+\!\Delta\sigma}
\end{align}
for some arbitrarily small constant $ \delta>0 $. The latter condition ensures that $ \sigma $ can be sufficiently varied by $ \Delta\sigma $ such that $ \calM_{\sigma\!+\!\Delta\sigma} $ varies over at least a $ \delta $-neighborhood of $ \calM_{\sigma} $. To allow for the evaluation of~\eqref{eq:local minimum condition b under D} under the partial knowledge on~$ b $, we naturally require for all $ \sigma\in\calP $ that $ \varepsilon_{\!\calM} $ in Assumption~\ref{ass:partial knowledge of a given CBF} is sufficiently large such that
\begin{align}
	\label{eq:vareps_M condition}
	D(\calM_{\sigma};\sigma\!\!+\!\!\Delta\sigma) \subseteq \calM\oplus\calB_{\varepsilon_{\!\calM}} \qquad \forall |\Delta\sigma|<\varepsilon.
\end{align}
Conditions~\eqref{eq:local minimum condition b under D}--\eqref{eq:vareps_M condition} are illustrated in Figure~\ref{fig:Delta sigma condition}. These conditions are sufficient to synthesize further CBFs based on the partially known CBF and the equivariance properties of the dynamics. To this end, the following theorem defines a subset of parameters $ \calS_{\calP} \subseteq \calP $ over which the maximum is taken, analogous to~\eqref{eq:CBF max definition}, but using $ \calS_{\calP} $ in place of $ \calP $. This yields a CBF based on the partial knowledge of~$ b $. 

\begin{theorem}
	\label{thm:outer approximation light saber}
	Let $ b: \bbR^{n}\rightarrow\bbR $ be a partially known CBF on a domain~$ \calD\subseteq\bbR^{n} $ as specified in Assumption~\ref{ass:partial knowledge of a given CBF}, and let dynamics~\eqref{eq:dynamics} be equivariant with respect to a diffeomorphism $ D(\cdot;\sigma): \bbR^{n} \rightarrow \bbR^{n} $ dependent on a scalar parameter $ \sigma\in\calP\subseteq\bbR $. Moreover, let conditions~\eqref{eq:local minimum condition b under D}--\eqref{eq:vareps_M condition} hold for any $ \sigma\in\calP $, and let the set-valued map $ \calS_{\calP} $ be defined as $ \calS_{\calP}(x) \coloneq \{ \sigma\in\calP \,|\, x\in\calM_{\sigma} \} $. Then, the value function $ B: \widehat{\calM} \rightarrow \bbR $ defined~as
	\begin{align}
		\label{eq:thm:outer approximation light saber 2}
		B(x) \coloneq \max_{\sigma\in \calS_{\calP}(x)} b_{\sigma}(x)
	\end{align}
	is a CBF in the Dini sense with respect to~\eqref{eq:dynamics} on a domain $ \calD_{B}\subseteq\widehat{\calM} $, if $ B $ is locally Lipschitz continuous on $ \calD_{B} $. The zero super-level set of $ b $ is given as 
	\begin{align}
		\label{eq:thm:outer approximation light saber 3}
		\calC = \bigcup_{\sigma\in\calP} \{ x\in\calM_{\sigma} \,|\, b_{\sigma}(x) \geq 0 \}.
	\end{align}
\end{theorem}

\begin{proof}
	As the Lipschitz continuity of $ B $ on its domain $ \calD_{B} $ is assumed for the moment --- we address it separately later on~---, we only need to show that~\eqref{eq:def cbf dini} holds for all $ x\in\calD_{B} $ to establish that $ B $ is a CBF in the Dini sense on $ \calD_{B} $. 
	
	To this end, let us consider the value of $ B $ at an arbitrary state $ x_{0}\in\calD_{B} $ given as $ B(x_{0}) = \max_{\sigma\in \calS_{\calP}(x_{0})} b(D(x_{0};\sigma)) $, and we define
	\begin{align}
		\label{seq:thm:outer approximation light saber aux 1c}
		\sigma_{0} &\coloneq \argmax_{\sigma\in \calS_{\calP}(x_{0})} b_{\sigma}(x_{0}).
	\end{align}
	If $ \sigma_{0} $ is not unique, we may choose any. Consider also $ \bm{\varphi}_{t} \coloneq \bm{\varphi}(t;x_{0},\bm{u}) $, a trajectory of the dynamic system~\eqref{eq:dynamics} starting at $ \bm{\varphi}_{0} \coloneq x_{0} $ and induced by a (sufficiently long) input trajectory $ \bm{u}\in\bm{\calU} $. In particular, we consider an input trajectory $ \bm{u} $ such that, for all $ t\geq0 $ close to 0, it holds
	\begin{align}
		\label{eq:thm:outer approximation light saber aux 2}
		db_{\sigma_{0}}(\bm{\varphi}_{t};f(\bm{\varphi}_{t},\bm{u}(t))) \geq -\alpha(b_{\sigma_{0}}(\bm{\varphi}_{t})).
	\end{align}
	Such $ \bm{u} $ exists since $ b_{\sigma_{0}} $ is a CBF by Lemma~\ref{lem:b_sigma is a CBF} with a domain including a $ \delta $-neighborhood of $ x_{0} $.
	
	We now investigate how the value of $ B $ changes along trajectory $ \bm{\varphi} $ close to time~$ 0 $. To this end, we compare the value of $ B(x_{0}) $ to $ B(\bm{\varphi}_{\Delta t}) $ for some $ \Delta t>0 $ such that $ \bm{\varphi}_{\Delta t}\in\calM_{\sigma_{0}}\oplus\calB_{\delta} $, with $ \delta $ as specified in condition~\eqref{eq:Delta sigma condition 2}. Such $ \Delta t $ clearly exists due to the continuity of $ \bm{\varphi} $. Analogously to~$ \sigma_{0} $, we define 
	\begin{align}
		\label{seq:thm:outer approximation light saber aux 1a}
		\sigma_{\Delta t} &\coloneq \argmax_{\sigma\in \calS_{\calP}(\bm{\varphi}_{\Delta t})} b_{\sigma}(\bm{\varphi}_{\Delta t})
	\end{align}
	and a suboptimal variant, restricted to the neighborhood of $ \sigma_{0} $, as
	\begin{align}
		\label{seq:thm:outer approximation light saber aux 1b}
		\tilde{\sigma}_{\Delta t} &\coloneq \argmax_{\sigma\in \calS_{\calP}(\bm{\varphi}_{\Delta t})\cap\calB_{\varepsilon}(\sigma_{0})} b_{\sigma}(\bm{\varphi}_{\Delta t}),
	\end{align}
	where $ \varepsilon=\varepsilon(\sigma_{0}) $ satisfies~\eqref{eq:Delta sigma condition 2} rendering the admissible set, from which $ \sigma $ is selected, non-empty. Defining $ \Delta\sigma \coloneq \sigma_{0} - \tilde{\sigma}_{\Delta t} $, where $ |\Delta\sigma|<\varepsilon $ by construction of $ \tilde{\sigma}_{\Delta t} $, we obtain 
	\begin{align*}
		B(\bm{\varphi}_{\Delta t}) &= b_{\sigma_{\Delta t}}(\bm{\varphi}_{\Delta t}) \stackrel{\eqref{seq:thm:outer approximation light saber aux 1b}}{\geq} b_{\tilde{\sigma}_{\Delta t}}(\bm{\varphi}_{\Delta t})\\ &\stackrel{\eqref{eq:local minimum condition b under D}}{\geq} b_{\tilde{\sigma}_{\Delta t}+\Delta\sigma}(\bm{\varphi}_{\Delta t}) = b_{\sigma_{0}}(\bm{\varphi}_{\Delta t}) \\ 
		&= b_{\sigma_{0}}(x_{0}) + \int_{0}^{\Delta t} db_{\sigma_{0}}(\bm{\varphi}_{t};f(\bm{\varphi}_{t},\bm{u}(t))) \, d\tau \\
		&\!\!\!\!\stackrel{\eqref{seq:thm:outer approximation light saber aux 1c},\eqref{eq:thm:outer approximation light saber aux 2}}{\geq} B(x_{0}) - \int_{0}^{\Delta t} \alpha(b_{\sigma_{0}}(\bm{\varphi}_{\tau})) \, d\tau
	\end{align*}
	and thus
	\begin{align}
		\label{eq:thm:outer approximation light saber aux 3}
		\frac{B(\bm{\varphi}_{\Delta t}) - B(x_{0})}{\Delta t} \geq -\frac{1}{\Delta t} \, \int_{0}^{\Delta t} \alpha(b_{\sigma_{0}}(\bm{\varphi}_{\tau})) \, d\tau.
	\end{align}
	Note that 
	\begin{subequations}
		\label{eq:thm:outer approximation light saber aux 4}
		\begin{align}
			&\liminf_{\Delta t\downarrow 0} \frac{B(\bm{\varphi}_{\Delta t}) - B(x_{0})}{\Delta t} \\
			\label{seq:thm:outer approximation light saber aux 4.2}
			\begin{split}
				&\qquad= \liminf_{\Delta t\downarrow 0} \frac{B(x_{0} + \Delta t f(x_{0},\bm{u}(0))) - B(x_{0})}{\Delta t} + \\
				&\qquad\quad\; \liminf_{\Delta t\downarrow 0} \frac{B(\bm{\varphi}_{\Delta t}) - B(x_{0} \!+\! \Delta t f(x_{0},\bm{u}(0)))}{\Delta t} 
			\end{split}\\
			&\qquad=\liminf_{\Delta t\downarrow 0} \frac{B(x_{0} + \Delta t f(x_{0},\bm{u}(0))) - B(x_{0})}{\Delta t} 
		\end{align}
	\end{subequations}
	where the second limit in~\eqref{seq:thm:outer approximation light saber aux 4.2} equals zero since $ \bm{\varphi}_{\Delta t} = \bm{\varphi}(\Delta t;x_{0},\bm{u}) = x_{0} + \Delta t f(x_{0},\bm{u}(0)) + \calO({\Delta t}^{2}) $ and therefore, by employing the Lipschitz constant of $ B $ denoted by $ L $, 
	\begin{align*}
		&\liminf_{\Delta t\downarrow0} \frac{|B(\bm{\varphi}(\Delta t;x_{0},\bm{u})) - B(x_{0} + \Delta t f(x_{0},\bm{u}(0)))|}{\Delta t} \\
		&= \liminf_{\Delta t\downarrow0} \frac{\splitfrac{|B(x_{0} + \Delta t f(x_{0},\bm{u}(0)) + \calO({\Delta t}^{2})) }{\hspace{2.5cm}- B(x_{0} + \Delta t f(x_{0},\bm{u}(0)))|}}{\Delta t} \\
		&\leq \liminf_{\Delta t\downarrow 0} \frac{L\calO({\Delta t}^{2})}{\Delta t} = 0.
	\end{align*}
	Then, by taking the limit on both sides of~\eqref{eq:thm:outer approximation light saber aux 3}, it follows
	\begin{align*}
		&dB(x_{0};f(x_{0},\bm{u}_{0})) = \liminf_{\Delta t\downarrow 0} \frac{B(x_{0} \!+\! \Delta t f\!(x_{0},\bm{u}\!(0))) - B(x_{0})}{\Delta t}  \\
		&\quad \stackrel{\eqref{eq:thm:outer approximation light saber aux 4}}{=} \liminf_{\Delta t\downarrow 0} \frac{B(\bm{\varphi}(\Delta t;x_{0},\bm{u})) - B(x_{0})}{\Delta t} \\
		&\quad\stackrel{\eqref{eq:thm:outer approximation light saber aux 3}}{\geq} -\alpha(b_{\sigma_{0}}(x_{0})) = -\alpha(B(x_{0})),
	\end{align*}
	and we conclude that~\eqref{eq:def cbf dini} holds. At last, we derive $ \calC $ as
	\begin{align*}
		\calC &= \{x \in\widehat{\calM} \, | \, B(x)\geq 0\} = \{x\in\widehat{\calM} \, | \, \max_{\sigma\in \calS_{\calP}(x)} b_{\sigma}(x) \geq 0\} \\
		&= \{x\in\widehat{\calM} \, | \, \exists \sigma\in \calS_{\calP}(x)\!: b_{\sigma}(x) \geq 0\}  \\
		&= \bigcup_{\sigma\in\calP} \{x\in\calM_{\sigma} \, | \, b_{\sigma}(x) \geq 0\},
	\end{align*}
	which concludes the proof.
\end{proof}

\begin{remark}
	Compared to Theorem~\ref{thm:outer approximation max cbf}, local condition~\eqref{eq:local minimum condition b under D} used in Theorem~\ref{thm:outer approximation light saber} allows for replacing the global maximization over all $ \sigma\in\calP $ in~\eqref{eq:CBF max definition} by a subset of parameters. The condition states that the value of $ b(D(x;\sigma)) $, $ x\in\calM_{\sigma} $, cannot be locally increased by an infinitesimal $ \Delta\sigma $-variation.
\end{remark}

The previous theorem assumed that $ B $ is locally Lipschitz continuous on a set $ \calD_{B}\subseteq\widehat{\calM} $. We now derive sufficient conditions under which this assumption holds. So far, $ D(\cdot;\sigma) $ was only required to be a diffeomorphism with respect to its first argument for any scalar parameter $ \sigma\in\calP $. However, to establish local Lipschitz continuity, stronger regularity properties of $ D $ are required. 

\begin{assumption}[Regularity Conditions for $ D $]
	\label{ass:regularity assumption on D}
	Let $ D $ and $ \calS_{\calP} $ satisfy the following regularity conditions:
	\begin{enumerate}[label=A\ref{ass:regularity assumption on D}.\arabic*, leftmargin=2.5em, labelwidth=2em, labelsep=0.5em]
		\item $ D(x;\cdot) $ is locally Lipschitz continuous in its second argument for every $ x\!\in\!\calD_{B} $;
		\label{ass:regularity assumption on D 1} 		
		\item $ \calS_{\calP} $ admits a Lipschitz continuous \emph{single-valued localization}; that is, for any $ x_{0}\in\calD_{B} $, $ \sigma_{0}\in \calS_{\calP}(x_{0}) $, there exist neighborhoods $ \calB_{\tilde{\delta}}(x_{0}) $ and $ \calB_{\tilde{\varepsilon}}(\sigma_{0}) $ of $ x_{0} $ and $ \sigma_{0} $, respectively, where $ \tilde{\delta},\tilde{\varepsilon}>0 $, such that the map 
		\begin{align}
			\label{eq:single valued localization S_P}
			x \mapsto \widetilde{\calS}_{\calP}^{\sigma_{0}}(x) \coloneq \calS_{\calP}(x) \cap \calB_{\tilde{\varepsilon}}(\sigma_{0})
		\end{align}
		is single-valued and Lipschitz continuous on $ \calB_{\tilde{\delta}}(x_{0}) $.
		\label{ass:regularity assumption on D3}
	\end{enumerate}
\end{assumption}

The latter of the assumptions is equivalent to the local existence of a Lipschitz continuous inverse of $ D $ with respect to its second argument. Single-valued localizations~\cite[p.~393ff.]{Rockafellar2009} can be seen as a generalization of local homeomorphisms for set-valued mappings. In particular,~\ref{ass:regularity assumption on D3} implies that for any $ \sigma_{0}\in\calP $, sets $ \calM_{\sigma'} $ with $ \sigma' $ in an arbitrarily small neighborhood of $ \sigma_{0} $ must be disjoint. Based on the additional regularity assumptions, the local Lipschitz continuity of~$ B $ can be established. 

\begin{lemma}
	\label{lemma:continuity of B in light saber theorem}
	Let the premises of Theorem~\ref{thm:outer approximation light saber} hold, and suppose $ D $ satisfies the regularity conditions in Assumption~\ref{ass:regularity assumption on D}. Then $ B $, as defined in~\eqref{eq:thm:outer approximation light saber 2}, is locally Lipschitz continuous on an open domain~$ \calD_{B}\subseteq\widehat{\calM} $, provided that for all $ x\in \calD_{B} $ 
	\begin{align}
		\label{eq:lemma:continuity of B}
		\exists \sigma\!\in\! \calS_{\calP}(x)\!: \, x\in\partial_{\varepsilon\,}\calM_{\sigma} \!\qquad\! \Rightarrow \qquad B(x)\!>\!b_{\sigma}(x),
	\end{align}
	where $ \partial_{\varepsilon\,}\calM_{\sigma} \!\coloneq\! \partial \calM^{\varepsilon}_{\sigma} \!\cap\! \calM_{\sigma} $ and $ \varepsilon\!\coloneq\!\varepsilon(\sigma) $. As a special case, if $ \calS_{\calP} $ maps each point in $ \calD_{B} $ to a singleton, then $ B $ is locally Lipschitz continuous on $ \calD_{B} $. 
\end{lemma}
\begin{proof}
	Consider $ B $ at any point $ x_{0}\in\calD_{B} $, and let $ \Sigma(x_{0}) $ denote the set of parameters $ \sigma_{0} $ that satisfy~\eqref{seq:thm:outer approximation light saber aux 1c}, or equivalently, for which $ B(x_{0})=b_{\sigma_{0}}(x_{0}) $. Using the fact that $ B(x_{0}) = b_{\sigma_{0}}(x_{0}) $ for any $ \sigma_{0}\in\Sigma(x_{0}) $, we conclude from~\eqref{eq:lemma:continuity of B} by contraposition that $ x_{0}\notin\partial_{\varepsilon(\sigma_{0})\,}\calM_{\sigma_{0}} $, and thus $ x_{0}\in\text{Int}(\calM_{\sigma_{0}}^{\varepsilon(\sigma_{0})})\cap\calM_{\sigma_{0}} $. Hence, $ x_{0}\in\bigcap_{\sigma_{0}\in\Sigma(x_{0})}\text{Int}(\calM_{\sigma_{0}}^{\varepsilon(\sigma_{0})}) $ and, as $ x_{0} $ lies in the interior of this intersection, there exists a sufficiently small constant $ \tilde{\delta}\in(0,\delta] $ with $ \delta $ as per~\eqref{eq:Delta sigma condition 2} such that $ \calB_{\tilde{\delta}}(x_{0})\subset\bigcap_{\sigma_{0}\in\Sigma(x_{0})}\text{Int}(\calM_{\sigma_{0}}^{\varepsilon(\sigma_{0})}) $. By~\ref{ass:regularity assumption on D3}, there exists a pair $ (\tilde{\varepsilon}, \tilde{\delta}) $ of sufficiently small positive constants, such that~\eqref{eq:single valued localization S_P} is a single-valued mapping on $ \calB_{\tilde{\delta}}(x_{0}) $ for some $ \tilde{\varepsilon}\leq\varepsilon(\sigma_{0}) $. 
	
	Consider now any $ x_{1}\in\calB_{\tilde{\delta}}(x_{0}) $ and define, analogously to before, the set $ \Sigma(x_{1}) $ as the set of parameters $ \sigma_{1} $ for which $ B(x_{1}) = b_{\sigma_{1}}(x_{1}) $. By definition, $  \Sigma(x_{0})$ and $ \Sigma(x_{1}) $ are subsets of $ \calS_{\calP}(x_{0}) $ and $ \calS_{\calP}(x_{1}) $, respectively. As shown before, $ \calS_{\calP} $ admits for $ \tilde{\varepsilon} $ and $ \tilde{\delta} $ a single-valued localization on $ \calB_{\tilde{\delta}}(x_{0}) $, and thus there exists a unique pair $ (\sigma_{0},\sigma_{1})\in\Sigma(x_{0})\times\Sigma(x_{1}) $ such that $ |\Delta\sigma|<\tilde{\varepsilon} $, where $ \Delta\sigma\coloneq\sigma_{1}-\sigma_{0} $. In particular, as $ \sigma_{1}\coloneq \sigma_{1}(x) \coloneq \widetilde{\calS}_{\calP}^{\sigma_{0}}(x_{1}) $, $ \Delta\sigma $ depends on $ x_{1} $ and we write $ \Delta\sigma(x_{1}) = \sigma_{0} - \widetilde{\calS}_{\calP}^{\sigma_{0}}(x_{1}) $. Due to the Lipschitz continuity of $ \widetilde{\calS}_{\calP}^{\sigma_{0}} $, also $ \Delta\sigma $ is Lipschitz continuous. For the sake of completeness, note also that $ \sigma_{0} \coloneq \widetilde{\calS}_{\calP}^{\sigma_{0}}(x_{0}) $.
	Let us now denote the local Lipschitz constants of $ b $, $ \Delta\sigma $, and $ D $ with respect to its first and second argument as $ L_{b} $, $ L_{\Delta\sigma} $, $ L_{D,x} $ and $ L_{D,\sigma} $, respectively. Then, defining $ \widetilde{B}_{\sigma_{0}}(x)\coloneq b_{\sigma_{1}}(x)\big|_{\sigma_{1}=\widetilde{\calS}_{\calP}^{\sigma_{0}}(x)} $ on $ \calB_{\tilde{\delta}}(x_{0}) $ for any $ \sigma_{0}\in\Sigma(x_{0}) $, we obtain its Lipschitz continuity at $ x_{0} $ as
	\begin{align*}
		&|\widetilde{B}_{\sigma_{0}}(x_{0}) \!-\! \widetilde{B}_{\sigma_{0}}(x_{1})| = |b(D(x_{0};\sigma_{0})) \!-\! b(D(x_{1};\sigma_{1}))| \\
		&\quad\leq L_{b} \, |D(x_{0};\sigma_{0}) \!-\! D(x_{1};\sigma_{0})| + L_{b} \, |D(x_{1};\sigma_{0}) \!-\! D(x_{1};\sigma_{1})| \\
		&\quad\leq L_{b} (L_{D,x} + L_{D,\sigma} L_{\Delta\sigma}) \, |x_{0}-x_{1}|. 
	\end{align*}
	Noting that for any $ x\in\calB_{\tilde{\delta}}(x_{0}) $
	\begin{align*}
		B(x) = \max_{\sigma\in \calS_{\calP}(x)} b_{\sigma}(x) = \max_{\sigma\in\Sigma(x)} b_{\sigma}(x) = \max_{\sigma\in\Sigma(x)} \widetilde{B}_{\sigma}(x),
	\end{align*}
	we conclude, together with the Lipschitz continuity of $ \widetilde{B}_{\sigma} $, the local Lipschitz continuity of $ B $ in $ x_{0} $ as the maximum operator preserves the Lipschitz continuity. Because the derivation holds for any $ x_{0}\in\calD_{B} $, we conclude that $ B $ is locally Lipschitz continuous on $ \calD_{B} $. In the case that $ \calS_{\calP} $ maps each point in $ \calD_{B} $ to singleton, then \eqref{eq:lemma:continuity of B} trivially holds for any open domain~$ \calD_{B} $ and the result follows from the general case.
\end{proof}

\begin{figure}[t]
	\centering
	\begin{subfigure}{0.53\columnwidth}
		\centering
		\def\svgwidth{1\columnwidth}
		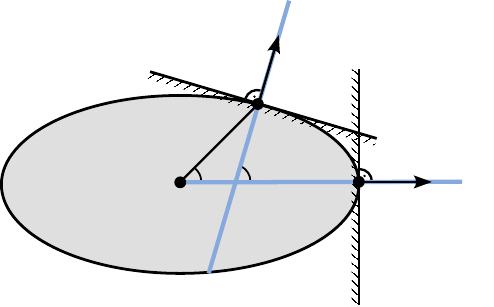
		\caption{}
		\label{fig:concept_light_saber_ellipse_small}
	\end{subfigure}
	\hfill
	\begin{subfigure}{0.45\columnwidth}
		\centering
		\def\svgwidth{1\columnwidth}
\begingroup%
  \makeatletter%
  \providecommand\color[2][]{%
    \errmessage{(Inkscape) Color is used for the text in Inkscape, but the package 'color.sty' is not loaded}%
    \renewcommand\color[2][]{}%
  }%
  \providecommand\transparent[1]{%
    \errmessage{(Inkscape) Transparency is used (non-zero) for the text in Inkscape, but the package 'transparent.sty' is not loaded}%
    \renewcommand\transparent[1]{}%
  }%
  \providecommand\rotatebox[2]{#2}%
  \newcommand*\fsize{\dimexpr\f@size pt\relax}%
  \newcommand*\lineheight[1]{\fontsize{\fsize}{#1\fsize}\selectfont}%
  \ifx\svgwidth\undefined%
    \setlength{\unitlength}{283.19109819bp}%
    \ifx\svgscale\undefined%
      \relax%
    \else%
      \setlength{\unitlength}{\unitlength * \real{\svgscale}}%
    \fi%
  \else%
    \setlength{\unitlength}{\svgwidth}%
  \fi%
  \global\let\svgwidth\undefined%
  \global\let\svgscale\undefined%
  \makeatother%
  \begin{picture}(1,0.58916084)%
    \lineheight{1}%
    \setlength\tabcolsep{0pt}%
    \put(0,0){\includegraphics[width=\unitlength,page=1]{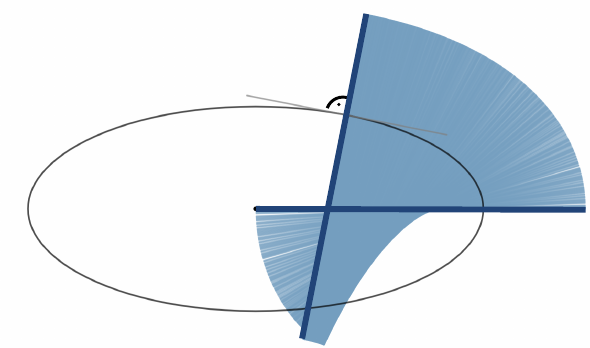}}%
    \put(0.58848147,0.5738817){\color[rgb]{0,0,0}\makebox(0,0)[lt]{\lineheight{1.25}\smash{\begin{tabular}[t]{l}$\calM_{\sigma}$\end{tabular}}}}%
    \put(0.91768319,0.16069683){\color[rgb]{0,0,0}\makebox(0,0)[lt]{\lineheight{1.25}\smash{\begin{tabular}[t]{l}$\calM$\end{tabular}}}}%
    \put(0,0){\includegraphics[width=\unitlength,page=2]{light_saber_method.pdf}}%
    \put(0.52601658,0.48080041){\color[rgb]{0,0,0}\makebox(0,0)[lt]{\lineheight{1.25}\smash{\begin{tabular}[t]{l}$\sigma$\end{tabular}}}}%
    \put(0,0){\includegraphics[width=\unitlength,page=3]{light_saber_method.pdf}}%
  \end{picture}%
\endgroup%

		\caption{}
		\label{fig:light_saber_method}
	\end{subfigure}
	\caption{Application of Theorem~\ref{thm:outer approximation light saber} to CBF synthesis. In the construction, $ \calM $ is shifted around the constraint set leaving the values of the partially known CBF $ b $ behind as a trace.}
	\label{fig:general_light_saber_method}
	\vspace{-\baselineskip}
\end{figure}

\begin{remark}
	The additional assumptions, namely Assumption~\ref{ass:regularity assumption on D} and~\eqref{eq:lemma:continuity of B}, invoked in the previous lemma can be intuitively understood from Figure~\ref{fig:light_saber_method}. It illustrates that $ D $ can be viewed as a transformation that moves set $ \calM $ along the boundary of a constraint as parameter $ \sigma $ changes. Assumption~\ref{ass:regularity assumption on D3} is fulfilled if $ \calM_{\sigma} $ and $ \calM_{\sigma\!+\!\Delta\sigma} $ do not intersect for arbitrarily small variations~$ \Delta\sigma $. In the case of larger variations $ \Delta\sigma $, however, the sets may overlap as indicted by the dark blue region in Figure~\ref{fig:light_saber_max}. Then~\eqref{eq:lemma:continuity of B} prevents a discontinuity on the boundary of the dark blue region. Note that \ref{ass:regularity assumption on D3} can be relaxed by replacing $ \calS_{\calP} $ with the restricted version $ \widehat{\calS}_{\calP}(x) = \calS_{\calP} \cap \Sigma(x) $; the proof of Lemma~\ref{lemma:continuity of B in light saber theorem} then still applies as \ref{ass:regularity assumption on D3} is only applied to points in $ \Sigma(x_{0}) $ and $ \Sigma(x_{1}) $. Then, $ \calM_{\sigma} $ and $ \calM_{\sigma\!+\!\Delta\sigma} $ may even intersect for small variations of $ \Delta\sigma $.
\end{remark}

\begin{remark}
	Let $ B_{1} $ and $ B_{2} $ be CBFs as defined in~\eqref{eq:CBF max definition} and~\eqref{eq:thm:outer approximation light saber 2}, respectively, for the same diffeomorphism $ D $ and parameter set $ \calP $. Then clearly $ B_{1}(x) \geq B_{2}(x) $, and thus $ B_{2} $ is more conservative than $ B_{1} $. For application, however, Theorem~\ref{thm:outer approximation light saber} is more advantageous as it only requires partial knowledge of a given CBF.
\end{remark}

\subsection{Application}

We now outline the application of Theorem~\ref{thm:outer approximation light saber} to the CBF synthesis for a given state constraint~\eqref{eq:state constraint}, leveraging partial knowledge of a CBF. Intuitively, as illustrated in Figure~\ref{fig:light_saber_method}, the method shifts~$ \calM $ along the constraint boundary, leaving behind the values of the partially known CBF~$ b $ as a trace, which collectively form the values of the CBF $ B $, which is to be synthesized. The application is structured into the following four steps (see Fig.~\ref{fig:general_light_saber_method} for an illustration):

\begin{enumerate}[
		label=\textbf{Step~\arabic*:},
		labelsep=0.5em,       
		leftmargin=1em,       
		labelwidth=0pt,       
		itemindent=-0.45em,     
		align=left            
	]
	\item Begin by approximating constraint function~$ h $ with a more conservative constraint function $ \tilde{h}:\bbR^{n}\rightarrow\bbR $ such that $ \tilde{h}(x) \leq h(x) $ for all $ x\in\bbR^{n} $, see Figure~\ref{fig:concept_light_saber_ellipse_small}. Function $ \tilde{h} $ is chosen such that it approximates (though possibly conservatively) the original constraint function~$ h $ even when ``shifted'' along the boundary of the original constraint set~$ \calH $. Next, choose $ \calM\subseteq\bbR^{n} $ and compute the numeric values of a CBF for constraint $ \tilde{h} $ on $ \calM\oplus\calB_{\varepsilon_{\!\calM}} $, for instance using~\eqref{eq:finite horizon construction H} with $ h $ replaced by~$ \tilde{h} $. We identify the CBF as the partially known CBF~$ b $ in Theorem~\ref{thm:outer approximation light saber}.
	
	\item Choose a locally Lipschitz continuous parameterization $ p:\calP\rightarrow\bbR^{n} $, $ \calP\subseteq\bbR $, of the boundary of constraint set~$ \calH $ and a diffeomorphism $ D(\cdot;\sigma) $, $ \sigma\in\calP $, such that dynamics~\eqref{eq:dynamics} are equivariant with respect to $ D $, and it holds for any $ \sigma\in\calP $, $ x\in\bbR^{n} $ that $ \tilde{h}(D(x;\sigma))\leq h(x) $, $ D(p(\sigma);\sigma) = p(0) $ and $ D(\calM;0) = \calM $. Thereby, diffeomorphism $ D $ shifts $ \tilde{h} $ along the boundary of $ \calH $ as depicted in Fig.~\ref{fig:concept_light_saber_ellipse_small}, and it holds $ \calC\subseteq\calH $ for the zero super-level set of $ b $ due to~\eqref{eq:thm:outer approximation light saber 3}.  
	
	\item Verify condition~\eqref{eq:local minimum condition b under D}; refine $ \calM $ or $ \tilde{h} $ if necessary.
	
	\item Construct $ B $ by using~\eqref{eq:thm:outer approximation light saber 2}. Intuitively, by varying parameter~$ \sigma $, $ \calM $ is shifted along the obstacle as shown in Fig.~\ref{fig:light_saber_method} and the values of the CBF on~$ \calM $ are left as a trace.
\end{enumerate}

 After following the previous steps, Theorem~\ref{thm:outer approximation light saber} can be applied and we conclude the following. 

\begin{corollary}
	Consider state constraint~\eqref{eq:state constraint}. If $ B $ is synthesized according to Steps~1-4 and is locally Lipschitz continuous, then $ B $ is a CBF in the Dini sense, satisfing~$ \calC\subseteq\calH $.
\end{corollary}

We note that selecting a set~$ \calM $ and a constraint function~$ \tilde{h} $ such that~$ b $ satisfies condition~\eqref{eq:local minimum condition b under D} may be non-trivial in some cases. However, once these are found, the method offers substantial computational advantages. In the subsequent examples, we apply the above steps to the synthesis of a CBF for the kinematic bicycle dynamics and non-trivial state constraints.  

\subsection{Examples}

The first example focuses on an elliptic obstacle. However, the same reasoning extends to any other smooth convex obstacle by modifying parameterization $ p $ of the boundary of~$ \calH $.

\begin{figure}
	\centering
		\begin{minipage}{0.45\columnwidth}
		\centering
		\vspace{0.5cm}
		\includegraphics[scale=0.16]{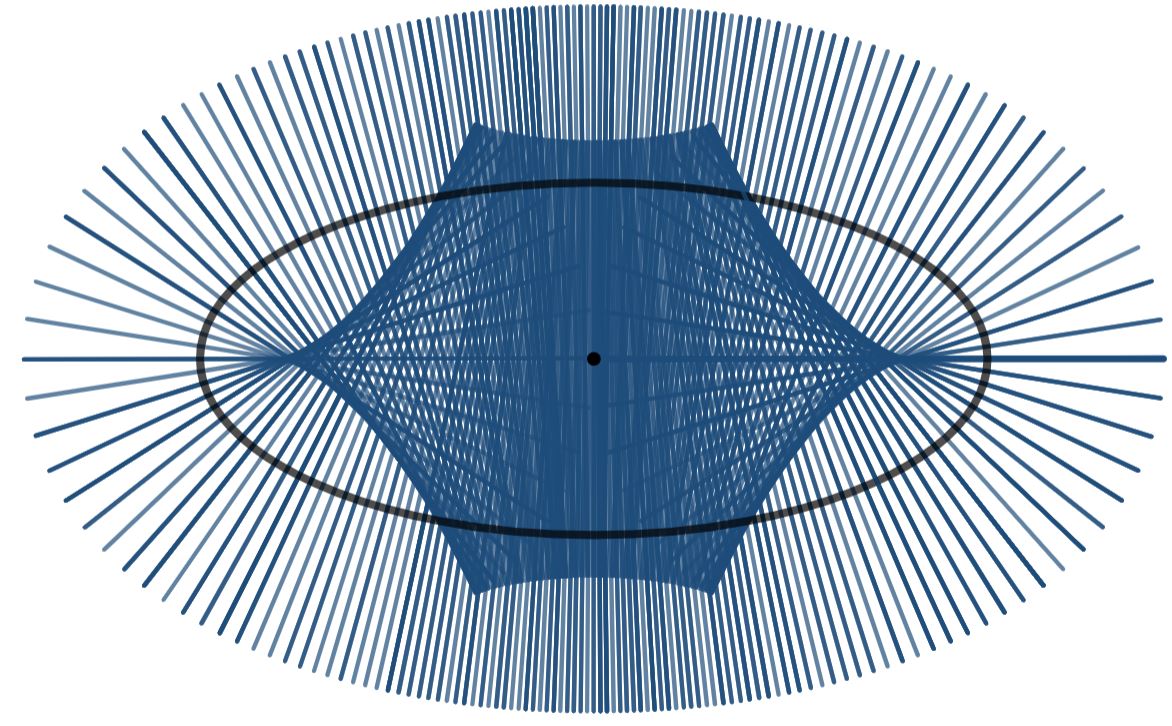}
		\caption{The dark blue region marks those states covered by $ \calM_{\sigma} $ for multiple $ \sigma $.}
		\label{fig:light_saber_max}
		\vspace{-\baselineskip}
	\end{minipage}
	\hfill
	\begin{minipage}{0.45\columnwidth}
		\centering
		\def\svgwidth{1\columnwidth}
		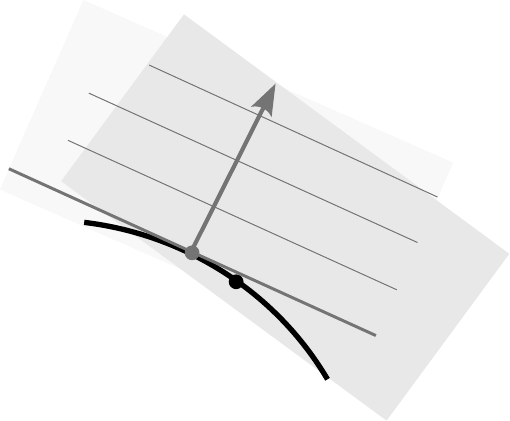
		\caption{Geometric construction for condition~\eqref{eq:local minimum condition b under D} in Example~\ref{exmp:bicycle model light saber}.}
		\label{fig:light_saber_convex_obstacle_condition_eval}
		\vspace{-\baselineskip}
	\end{minipage}%
\end{figure}

\begin{example}
	\label{exmp:bicycle model light saber}
	Let us revisit the kinematic bicycle model~\eqref{eq:bicycle model} and consider an elliptical obstacle as in Fig.~\ref{fig:concept_light_saber_ellipse_small}. Accordingly, any point in the set to be rendered invariant must satisfy $ h(\mathbf{x}) \coloneq \frac{x^2}{a^2} + \frac{y^2}{b^2} - 1 \geq 0 $ for some constants $ a,b>0 $. Since no diffeomorphism~$ D $ exists under which both $ h $ is symmetric and the bicycle kinematics are equivariant, we resort to the results developed in Theorem~\ref{thm:outer approximation light saber}. Our objective is to construct a CBF~$ B $, whose zero super-level set satisfies~$ \calC\subseteq\calH $. 
	
	\textbf{Step~1:} We approximate constraint function~$ h $ by $ \tilde{h}(\mathbf{x})\coloneq\langle [1,0]^{T}, \mathbf{x}_{\text{pos}} - [a,0]^{T} \rangle \leq h(\mathbf{x}) $ as depicted in Fig.~\ref{fig:concept_light_saber_ellipse_small}. Furthermore, we choose $ \calM\coloneq \{ \mathbf{x} \in \bbR^{3} \, | \, x\in[0,a+\Delta a], \; y = 0, \; \psi = [0,2\pi] \} $, $ \Delta a>0 $, and compute $ b $ on $ \calM\oplus\calB_{\varepsilon_{\!\calM}} $ for any $ \varepsilon_{\!\calM}>0 $ by using~\eqref{eq:finite horizon construction H} with $ h $ replaced by~$ \tilde{h} $. Thereby, the values of $ b $ on $ \calM\oplus\calB_{\varepsilon_{\!\calM}} $ correspond to those of $ H_{T} $.
	
	\textbf{Step~2:} We parameterize the boundary of the ellipse by $ p(\sigma) = \begin{bsmallmatrix}
		a \cos(\sigma) \\ b \sin(\sigma)
	\end{bsmallmatrix} $, where $ \sigma\in\calP\coloneq [0,2\pi) $. We then define 
	\begin{align}
		\label{eq:diffeo bicycle light saber}
		D(\mathbf{x};\sigma) \coloneq \begin{bsmallmatrix}
			R(-\phi(\sigma))(\mathbf{x}_{\text{pos}} - p(\sigma_{0}+\sigma)) + p(\sigma_{0}) \\ \psi + \phi(\sigma_{0}) - \phi(\sigma)
		\end{bsmallmatrix},
	\end{align}
	where $ \sigma_{0} $ is a constant such that $ D(\calM;\sigma_{0}) = \calM $ (by convention: $ \sigma_{0} = 0 $); function $ \phi(\sigma)\coloneq\arccos\!\left(\!\frac{n(\sigma_{0}) n(\sigma)}{||n(\sigma_{0})|| \, ||n(\sigma)||}\!\right) $ is defined as the angle between the normal vectors $ n(\sigma_{0}) $ and $ n(\sigma) $ in $ p(\sigma_{0}) $ and $ p(\sigma) $, respectively, given by $ n(\sigma) \coloneq \begin{bsmallmatrix}
		b \cos(\sigma) \\ -a \sin(\sigma)
	\end{bsmallmatrix} $. Intuitively, transformation $ D $ shifts $ \tilde{h} $ tangentially along the boundary of the ellipse and thus $ \tilde{h}(D(\mathbf{x};\sigma))\leq h(\mathbf{x}) $ for any $ \mathbf{x}\in\bbR^{n} $, $ \sigma\in\calP $. Consequently, $ \calC\subseteq\calH $ by~\eqref{eq:thm:outer approximation light saber 3}. For an illustration refer to Fig.~\ref{fig:concept_light_saber_ellipse_small}. As $ D $ is a composition of translations and rotations --- transformations under which the kinematic bicycle model has been shown to be equivariant (see Section~\ref{subsec:exampel Kinematic bicycle model}) --- the equivariance of~\eqref{eq:bicycle model} under $ D $ follows.

	\textbf{Step~3:} From the geometric construction in Figure~\ref{fig:light_saber_convex_obstacle_condition_eval}, which utilizes the symmetry properties of $ b $, inherited from those of the constraint function $ \tilde{h} $ and the translational equivariance of the bicycle kinematics (Section~\ref{subsec:exampel Kinematic bicycle model}), the satisfaction of condition~\eqref{eq:local minimum condition b under D} follows directly. 
	
	\textbf{Step~4:} At last, since all premises of Theorem~\ref{thm:outer approximation light saber} are satisfied, we conclude that $ B $, defined in~\eqref{eq:thm:outer approximation light saber 2}, is a CBF. As $ D $ is smooth in both of its arguments, the regularity assumption, Assumption~\ref{ass:regularity assumption on D}, holds. Moreover, as also~\eqref{eq:lemma:continuity of B} holds, we conclude the Local Lipschitz continuity of~$ B $~by Lemma~\ref{lemma:continuity of B in light saber theorem}, even though $ \calM $ overlaps as it is shifted along the constraint boundary (see dark blue region in Figure~\ref{fig:light_saber_max}). 
\end{example}

The construction of a CBF for an elliptical obstacle extends analogously to any smooth, convex obstacle. Only the parameterization $ p $ and the definition of the angle~$ \phi $ need to be adjusted. With a minor modification, even non-smooth obstacles --- such as those with a corner --- can be handled. 

\begin{figure}
	\centering
	\def\svgwidth{0.25\columnwidth}
\begingroup%
  \makeatletter%
  \providecommand\color[2][]{%
    \errmessage{(Inkscape) Color is used for the text in Inkscape, but the package 'color.sty' is not loaded}%
    \renewcommand\color[2][]{}%
  }%
  \providecommand\transparent[1]{%
    \errmessage{(Inkscape) Transparency is used (non-zero) for the text in Inkscape, but the package 'transparent.sty' is not loaded}%
    \renewcommand\transparent[1]{}%
  }%
  \providecommand\rotatebox[2]{#2}%
  \newcommand*\fsize{\dimexpr\f@size pt\relax}%
  \newcommand*\lineheight[1]{\fontsize{\fsize}{#1\fsize}\selectfont}%
  \ifx\svgwidth\undefined%
    \setlength{\unitlength}{67.42249556bp}%
    \ifx\svgscale\undefined%
      \relax%
    \else%
      \setlength{\unitlength}{\unitlength * \real{\svgscale}}%
    \fi%
  \else%
    \setlength{\unitlength}{\svgwidth}%
  \fi%
  \global\let\svgwidth\undefined%
  \global\let\svgscale\undefined%
  \makeatother%
  \begin{picture}(1,0.96660586)%
    \lineheight{1}%
    \setlength\tabcolsep{0pt}%
    \put(0,0){\includegraphics[width=\unitlength,page=1]{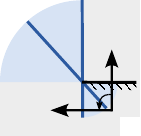}}%
    \put(0.81648214,0.20788895){\color[rgb]{0,0,0}\makebox(0,0)[lt]{\lineheight{1.25}\smash{\begin{tabular}[t]{l}\footnotesize{$\sigma_1$}\end{tabular}}}}%
    \put(0.87403854,0.46073506){\color[rgb]{0,0,0}\makebox(0,0)[lt]{\lineheight{1.25}\smash{\begin{tabular}[t]{l}\footnotesize{$n_1$}\end{tabular}}}}%
    \put(0.42706581,0.30215057){\color[rgb]{0,0,0}\makebox(0,0)[lt]{\lineheight{1.25}\smash{\begin{tabular}[t]{l}\footnotesize{$p_0$}\end{tabular}}}}%
    \put(0.34009289,0.06861823){\color[rgb]{0,0,0}\makebox(0,0)[lt]{\lineheight{1.25}\smash{\begin{tabular}[t]{l}\footnotesize{$n_2$}\end{tabular}}}}%
    \put(0.61265021,0.82094338){\color[rgb]{0,0,0}\makebox(0,0)[lt]{\lineheight{1.25}\smash{\begin{tabular}[t]{l}\footnotesize{$\calM$}\end{tabular}}}}%
    \put(0.26153039,0.77422107){\color[rgb]{0,0,0}\makebox(0,0)[lt]{\lineheight{1.25}\smash{\begin{tabular}[t]{l}\footnotesize{$\calM_{\sigma}$}\end{tabular}}}}%
    \put(0,0){\includegraphics[width=\unitlength,page=2]{corner_light_saber.pdf}}%
  \end{picture}%
\endgroup%

	\caption{Construction for a corner.}
	\label{fig:corner_light_saber}
	\vspace{-\baselineskip}
\end{figure}

\begin{example}
	Let us consider the kinematic bicycle model once more, but now for a constraint describing a corner, formally 
	\begin{align*}
		h(\mathbf{x}) \coloneq \max\{ \underbrace{\langle n_{1}, \mathbf{x}_{\text{pos}} \rangle  + c_{1}}_{\eqcolon h_{1}(\mathbf{x})}, \, \underbrace{\langle n_{2}, \mathbf{x}_{\text{pos}} \rangle + c_{2}}_{\eqcolon h_{2}(\mathbf{x})} \} \geq 0,
	\end{align*}
where $ c_{1}, c_{2} \in \bbR $ are some constants and linearly independent $ n_{1}, n_{2} \in\bbR^{2} $ such that the obstacle is convex as illustrated in Figure~\ref{fig:corner_light_saber}. Such constraint function is non-smooth. To construct a CBF, we define $ \tilde{h} \coloneq h_{1} $, the point $ p_{0}\in\bbR^{2} $ such that $ \langle n_{1} - n_{2}, p_{0} \rangle + (c_{1} - c_{2}) = 0 $, and $ \calM \coloneq \{ \mathbf{x}\in\bbR^{3} \, | \, \mathbf{x}_{\text{pos}} = p_{0} + \nu n_{1}, \; \nu\in\bbR \} $. Moreover, we choose the diffeomorphism~$ D $ to be the same as in~\eqref{eq:diffeo bicycle light saber}, however with $ p(\cdot) \equiv p_{0} $, $ \phi(\sigma) \coloneq \sigma $, $ \sigma_{0} = 0 $ and $ \calP \coloneq [\min\{\sigma_{0}, \sigma_{1}\}, \max\{\sigma_{0}, \sigma_{1}\}] $; $ \sigma_{1} $ is as indicated in Fig.~\ref{fig:corner_light_saber}. Given this, condition~\eqref{eq:local minimum condition b under D} is verified analogously to the previous example. With this, $ B $ defined in~\eqref{eq:thm:outer approximation light saber 2} is a CBF for the corner. 
\end{example}

\vspace{-0.5\baselineskip}

\section{Numeric Examples}
\label{sec:numeric examples}

This section provides a numerical study of the computational gains achieved by leveraging equivariances and symmetries, supporting the theoretical findings empirically.

\vspace{-0.5\baselineskip}

\subsection{Performance Analysis}

\begin{table*}[t]
	\centering
	\begin{tabular}{l!{\vrule}cc!{\vrule}ccc!{\vrule}ccc!{\vrule}c!{\vrule}c}
		\toprule
		\multicolumn{1}{c|}{} & \multicolumn{2}{c|}{\textit{Specifications}} &  \multicolumn{3}{c|}{\textit{\#points computed}} & \multicolumn{5}{c}{\textit{Computation times (equivariance-based vs. direct)}} \\
		&  &  &  &  &  &  \multicolumn{3}{c|}{\textbf{equivariance-based synthesis}} & \multirow{2}{*}{\shortstack{\textbf{direct \cite{Wiltz2025b}}\\\text{[h:mm:ss]}}} & \multirow{2}{*}{\shortstack{\textbf{total e.-b./}\\\textbf{direct}}} \\
		& \textbf{domain} & \textbf{$ \calM $} & \textbf{explicit} & \textbf{total} & \textbf{ratio} & \textbf{explicit} & \textbf{inferred} & \textbf{total} &  & 
		\\
		\midrule
		single integrator & $ [-10,10]^{2} $ & \scriptsize$ [-15,0.1]\!\times\!\{0\} $ & 30 & 1681 & $ 1.78\% $ & 10.01s & 0.05s & 10.06s & $ 0\!:\!01\!:\!24 $ & $ 11.98\% $ \\
		single i. ($ v_{x,\text{min}} > 0 $) & $ [-10,10]^{2} $ & \scriptsize$ [-15,0.1]\!\times\!\{0\} $ & 30 & 1681 & $ 1.78\% $ & 12.65 & 0.05s & 12.70s & $ 0\!:\!01\!:\!11 $ & $ 17.89\% $ \\
		double integrator & \makecell[c]{
			\tiny\linespread{0.8}\selectfont
			\kern-0.4em$[-14,14]^2$\\[-4pt]\tiny
			\kern-0.4em$\times [-2.5,2.5]^2$
		} & \makecell[c]{
			\tiny\linespread{0.8}\selectfont
			\kern-0.4em$[-20,0.1]\!\times\!\{0\}$\\[-4pt]\tiny
			\kern-0.4em$\times [-2.5,2.5]^2$
		} & 4725 & 189 225 & $ 2.50\% $ & 318.62s & 18.74s & 337.36s & $ 2\!:\!49\!:\!46 $ & $ 3.31\% $ \\
		bicycle (less agile) & \makecell[c]{
			\tiny\linespread{0.8}\selectfont
			\kern-0.4em$[-15,15]^{2}$\\[-4pt]\tiny
			\kern-0.4em$\times [-\pi,\pi]$
		} & \makecell[c]{
			\tiny\linespread{0.8}\selectfont
			\kern-0.4em$[-21.3,0.1]$\\[-4pt]\tiny
			\kern-0.4em$\times\{0\}\times [-\pi,0]$
		} & 903 & 152 561 & $ 0.59\% $ & 171.22s & 10.46s & 181.68s & $ 5\!:\!01\!:\!42 $ & $ 1.00\% $ \\
		bicycle (more agile) & \makecell[c]{
			\tiny\linespread{0.8}\selectfont
			\kern-0.4em$[-10,10]^{2}$\\[-4pt]\tiny
			\kern-0.4em$\times [-\pi,\pi]$
		} & \makecell[c]{
			\tiny\linespread{0.8}\selectfont
			\kern-0.4em$[-14.2,0.1]$\\[-4pt]\tiny
			\kern-0.4em$\times\{0\}\times [-\pi,0]$
		}  & 630 & 68 921 & $ 0.91\% $ & 110.34s & 4.5s & 114.84s & $ 2\!:\!33\!:\!33 $ & $ 1.25\% $ \\
		unicycle & \makecell[c]{
			\tiny\linespread{0.8}\selectfont
			\kern-0.4em$[-10,10]^{2}$\\[-4pt]\tiny
			\kern-0.4em$\times [-\pi,\pi]$
		} & \makecell[c]{
			\tiny\linespread{0.8}\selectfont
			\kern-0.4em$[-14.2,0.1]$\\[-4pt]\tiny
			\kern-0.4em$\times\{0\}\times [-\pi,0]$
		} & 630 & 68 921 & $ 0.91\% $ & 69.81s & 4.18s & 73.99s & $ 1\!:\!38\!:\!12 $ & $ 1.26\% $ \\
		\bottomrule
	\end{tabular}
	\caption{Number of explicitly computed vs. the total number of computed points in the equivariance-based CBF synthesis, and comparison of the computation times -- split up for our equivariance-based method into the time for explicitly computed points, points inferred from equivariances and the total computation time --  with those of the direct method in~\cite{Wiltz2025b}.}
	\label{tab:computation time}
\end{table*}

\begin{figure*}
	\centering
	\begin{subfigure}{0.19\linewidth}
		\centering
		\includegraphics[width=1.0\linewidth]{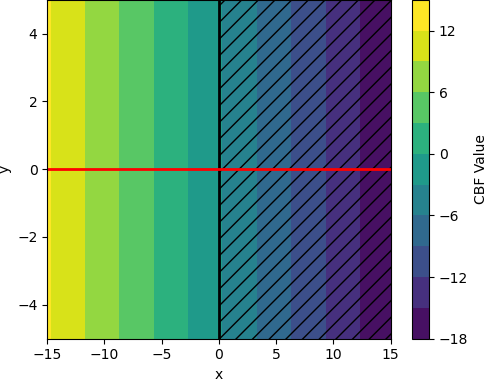}
		\caption{Line}
		\label{fig:line}
	\end{subfigure}%
	\hfill
	\begin{subfigure}{0.19\linewidth}
		\centering
		\includegraphics[width=1.0\linewidth]{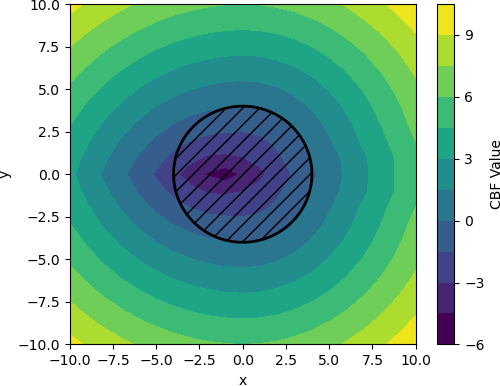}
		\caption{Circle}
		\label{fig:circle}
	\end{subfigure}
	\hfill
	\begin{subfigure}{0.19\linewidth}
		\centering
		\includegraphics[width=1.0\linewidth]{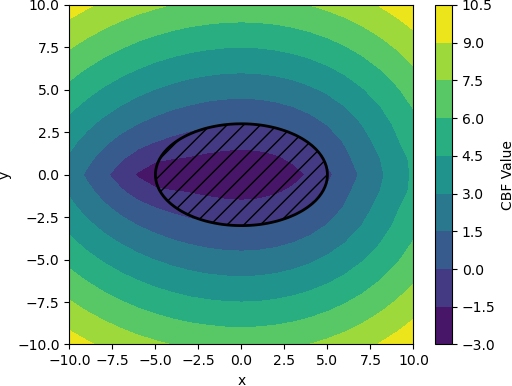}
		\caption{Ellipse}
		\label{fig:ellipse}
	\end{subfigure}
	\hfill
	\begin{subfigure}{0.19\linewidth}
		\centering
		\includegraphics[width=1.02\linewidth]{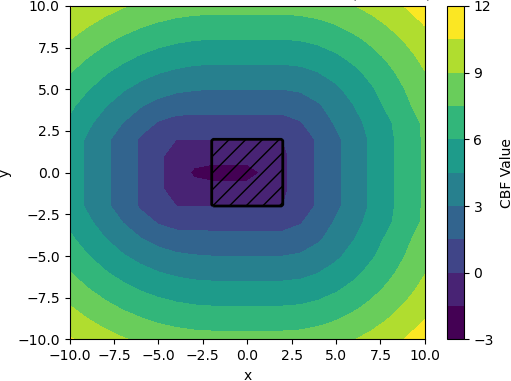}
		\caption{Square}
		\label{fig:square}
	\end{subfigure}
	\hfill
	\begin{subfigure}{0.19\linewidth}
		\centering
		\includegraphics[width=1.0\linewidth]{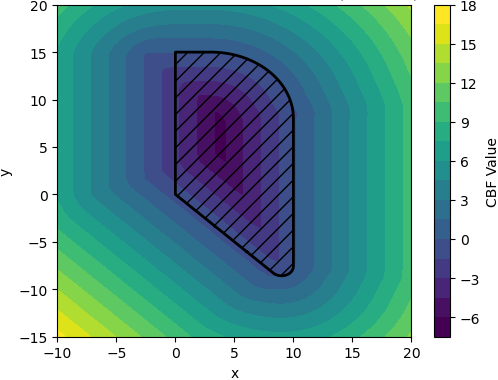}
		\caption{Convex obstacle}
		\label{fig:convex_obstacle}
	\end{subfigure}
	\caption{CBFs at orientation $\psi = 0$ for various convex obstacles, computed for the bicycle model ($v_{\text{min}} > 0$, $\zeta_{\text{max}} = 20\pi/180$). The red line in (a) marks the explicitly computed values from which all other CBFs are derived.}
	\label{fig:convex_obstacle_series}
	\vspace{-\baselineskip}
\end{figure*}

\begin{figure}
	\centering
	\begin{subfigure}{0.49\columnwidth}
		\centering
		\includegraphics[width=1.0\linewidth]{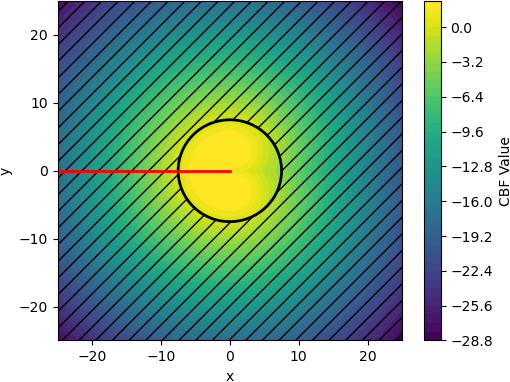}
		\caption{Circle}
		\label{fig:inner_circle}
	\end{subfigure}%
	\hfill
	\begin{subfigure}{0.49\columnwidth}
		\centering
		\includegraphics[width=0.98\linewidth]{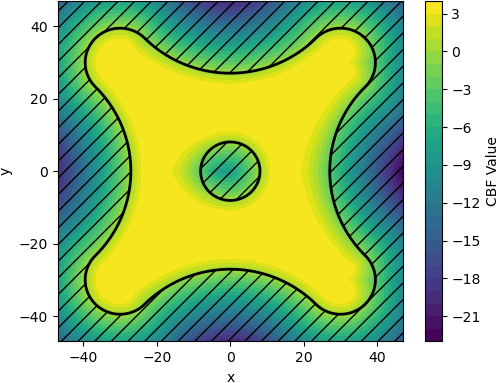}
		\caption{Non-convex obstacle}
		\label{fig:nonconvex_obstacle}
	\end{subfigure}
	\caption{CBFs at orientation $\psi = 0$ for non-convex obstacles, constructed based on the explicitly computed values marked by the red line in (a).}
	\label{fig:nonconvex_obstacle_series}
	\vspace{-\baselineskip}
\end{figure}

Leveraging equivariance and symmetry properties in the CBF synthesis offers substantial computational benefits, as indicated by our theoretical results. To support these findings numerically, we revisit the CBFs computed in our earlier work~\cite[Sec.~VII]{Wiltz2025b} and compare their computation times to those obtained when leveraging equivariances. The constraint function under consideration is 
\begin{align}
	\label{eq:circular constraint}
	h(x) = \sqrt{(x-x_{c})^{2} + (y-y_{c})^{2}} - r,
\end{align}
which defines a circular obstacle centered at $ (x_{c},y_{c}) $ with radius $ r $. We synthesize CBFs for three different types of dynamic systems:
\begin{enumerate}[label=\Alph*.]
	\item \emph{Single and double integrators} ($ \dot{x} = u $ and $ \ddot{x} = u $) as examples for basic first- and second-order linear systems. In Section~\ref{subsec:linear system}, we showed that the double integrator is equivariant under the rotation of position and velocity; an analogous result follows for the single integrator as a degenerate case. 
	\item The \emph{bicycle model} in~\eqref{eq:bicycle model} with minimum velocity $ v_{\text{min}}\!>\!0 $ as an example for a system that is not locally controllable. In the equivariance-based CBF synthesis, we exploit its rotational and orientational equivariance (see Section~\ref{subsec:exampel Kinematic bicycle model}). We consider a more and a less agile version with larger/smaller maximal steering angles $ |\zeta| \!<\! \zeta_{\text{max}} $. 
	\item The \emph{unicycle} as an example for a non-holonomic system; equivariances are analogous to those of the bicycle model. 
\end{enumerate}
System parameters, input constraints and the design parameters of the CBF are chosen to be the same as in~\cite{Wiltz2025b}. For the equivariance based computation of the CBF, we proceed as follows: we first explicitly compute the CBF on a grid defined on set $ \calM $ with respect to~\eqref{eq:circular constraint} by solving~\eqref{eq:finite horizon construction H} in each point (the number of explicitly computed points and the required computation time can be seen from Table~\ref{tab:computation time}); secondly, the values of all remaining points are obtained by exploiting equivariances as by~\eqref{eq:thm:outer approximation light saber 2}. Table~\ref{tab:computation time} compares the number of points explicitly computed and those computed based on equivariances with each other, as well as the computation time between computing the CBF values by solving~\eqref{eq:finite horizon construction H} in each point (direct application of~\cite{Wiltz2025b}) and a synthesis based on equivariances. By Theorem~\ref{thm:outer approximation light saber}, each of the computed functions is indeed a CBF. Since the bicycle model, as shown in Section~\ref{subsec:exampel Kinematic bicycle model}, is strongly equivariant, it follows by Theorem~\ref{thm:symmetry H_T} that the CBF directly computed based on~\cite{Wiltz2025b} and the CBF computed based on equivariances are identical. This is confirmed by comparing the numeric values of the computed CBFs\footnote{mean deviation 0.0031 (0.0012), maximal deviation 0.018 (0.017) for the less (more) agile bicycle model}. Therefore, we refer for a visualization of the computed CBFs and simulation results to~\cite{Wiltz2025b}. The Python code and the computation results to the examples presented in this section are available on Github\footnote{\url{https://github.com/KTH-DHSG/Equivariance_Based_CBF_Synthesis.git}}. The computations have been conducted on a 12th Gen Intel Core i9-12900K with 64GB RAM. The implementation uses the optimal control package Casadi~\cite{Andersson2019}.

\subsection{Synthesizing Multiple CBFs from a Single One}

Equivariances allowed us to reduce the computation times in the previous examples by up to $ 99\% $. Yet, the computation of the explicitly computed points takes by far the largest share of the overall computation time. This can be further improved by means of Theorem~\ref{thm:outer approximation light saber}. It allows us to derive CBFs for various, not necessarily symmetric constraints, while only requiring knowledge on a partially computed CBF. Thereby, already explicitly computed CBF values can be leveraged to the synthesis of CBFs for a multitude of constraints. To illustrate this, let us reconsider the bicycle model with input constraints $v_{\text{min}} > 0$, $\zeta_{\text{max}} = 20\pi/180$ (``less agile''). For synthesizing a CBF for any convex obstacle as depicted in Figure~\ref{fig:convex_obstacle_series}, it is sufficient to explicitly compute the CBF with respect to the constraint depicted in Figure~\ref{fig:line} on $ \calM=\{[x,y,\psi]^{T}\!\in\!\bbR^{3} \; | \; |x|\!\leq\! x_{\text{max}},\, y\!=\!0,\, \psi\!\in[-\pi,0]\} $ based on~\eqref{eq:finite horizon construction H}. All further values can be then inferred via equivariances. The same method is applicable also to non-convex constraints, see Figure~\ref{fig:nonconvex_obstacle_series}.


\section{Conclusion}
\label{sec:conclusion}

This paper investigated how equivariant dynamics and symmetric constraints can be leveraged for CBF synthesis. First, we identified the conditions under which these properties extend to a class of reachability-based CBFs. Since constraints are often only locally symmetric, or not symmetric at all, under the diffeomorphism that defines equivariance, we developed a more general framework that uses equivariances to synthesize CBFs in these cases. This approach requires only partial knowledge of another CBF. Numerous examples supported the theory and illustrated its practical usefulness and the broad applicability of equivariance-based CBF synthesis.


\balance

\bibliographystyle{IEEEtran}
\bibliography{/Users/wiltz/CloudStation/JabBib/Research/000_MyLibrary}


\end{document}